\documentclass[aps,pre,10pt,twocolumn,superscriptaddress,nonotesinbibliography]{revtex4-1}
\usepackage{graphicx}
\usepackage{amsmath}
\usepackage{amsfonts}
\usepackage{amssymb}
\usepackage{color}

\DeclareMathOperator{\upd}{d\!}

\def\wid{0.45}

\begin{document}

\title{Memory Effects in Active Particles with Exponentially Correlated Propulsion}

\author{Cato Sandford}
\affiliation{Department of Physics and Center for Soft Matter Research, New York University, 726 Broadway, New York, NY 10003, USA}
\author{Alexander Y. Grosberg}
\affiliation{Department of Physics and Center for Soft Matter Research, New York University, 726 Broadway, New York, NY 10003, USA}
\date{\today}

\begin{abstract}
The Ornstein--Uhlenbeck Particle (OUP) model imagines a microscopic swimmer propelled by an active force which is correlated with itself on a finite time-scale. Here we investigate the influence of external potentials on an ideal suspension of OUPs, in both one and two spatial dimensions, with particular attention paid to the pressure exerted on ``confining walls''. We employ a mathematical connection between the local density of OUPs and the statistics of their propulsion force to demonstrate the existence of an equation of state in one dimension. In higher dimensions we show that active particles generate a non-conservative force field in the surrounding medium. A simplified far-from-equilibrium model is proposed to account for OUP behaviour in the vicinity of potentials. Building on this, we interpret simulations of OUPs in more complicated situations involving asymmetrical and spatially curved potentials, characterising the inhomogeneous local stresses which result in terms of competing active length-scales.
\end{abstract}

\maketitle

\section{Introduction}
\label{sec:intro}

Active swimmers \cite{Mar++13} are particles which have access to an external or internal source of energy, and continually dissipate that energy to perform persistent motion. This breaks detailed balance, and gives rise to steady-state currents in phase space and real space, as well as non-Gibbsian steady state distributions \cite{Lee13,E+G13,YMM14}. A range of other macroscopic non-equilibrium phenomena have been investigated, including wall-dependent pressures \cite{Sol++15}, jamming and phase separation for interacting swimmers \cite{T+C08,C+T15,Mar++16,YMM14,Sol++15--P-decomposition}, and novel first-passage properties \cite{HDL89} for non-interacting swimmers.

Two models of microscopic swimmer dynamics have received the lion's share of attention. Active Brownian particles (ABPs) and run-and-tumble particles (RTPs) both exert a propulsion force of a constant magnitude and randomly fluctuating direction. For ABPs the orientation vector undergoes gradual rotational diffusion, while for RTPs it is instantaneous randomised at Poissonian time intervals \cite{C+T15}.
A third model, which will be the focus of this article, is the Ornstein--Uhlenbeck particle (OUP). OUPs' propulsion force $\vec \eta$ varies randomly in both magnitude and direction, with each component being exponentially self-correlated on a time-scale $\tau$.  Although previous studies \cite{HDL89,DHL87,Luc05,J+H87,KML14,M+M15,Mag++15,CN1} have have sought to characterise the OUP model in a variety of situations, it remains less thoroughly explored than the other two.

The variable propulsion force of the OUP model presents mathematical challenges not encountered in the other models; but it also gives rise to some interesting physics which we explore below.
Of particular interest to us in this regard is the local pressure exerted by OUPs on confining walls. Elsewhere \cite{CN1} we have investigated non-equilibrium mechanical phenomena which arise from the combination of OUPs' spatial correlations and their interactions with potentials;
here we present a complementary (and more thorough) discussion, and note a number of new effects.
As in our previous work, we do not treat the correlated propulsion force $\vec \eta$ as a nuisance to be eliminated (as others have done, e.g. \cite{J+H87}): instead we consider the full $(\vec x,\vec \eta)$ phase space, which turns out to be highly informative and helps us to develop physical insights.
(Since we concentrate on somewhat complex geometries and do not assume the system to be close to equilibrium, the range of analytical progress is often limited, and so many of our conclusions in latter sections of the paper will be based on numerical simulations.)

The paper is organised as follows. In section~\ref{sec:model} we introduce the OUP model and the notations which will be used throughout. Since a general solution for the steady state density is elusive, we investigate moments of the distribution in section~\ref{sec:pressure} and connect them with the mechanical pressure. In section~\ref{sec:GO} we introduce a simplified model which gives qualitive insight into how OUPs' propulsion force is affected by the proximity of potentials. We review in section~\ref{sec:CAR_DL} simulations of OUPs in a simple symmetrical 1D potential, to build intuition before considering asymmetrical potentials which develop net forces forbidden in ideal equilibrium systems. A one-dimensional potential is invesitgated in section~\ref{sec:CAR_ML}, while sections~\ref{sec:CAR_UL} and~\ref{sec:polar} concern 2D situations which permit the existence of geometrical curvature.
We conclude in section~\ref{sec:conclusion}.  Several technical developments are relegated to appendices.

\section{Swimmers with Correlated Propulsion}
\label{sec:model}

In one spatial dimension, the overdamped microscopic dynamics for the OUP position $x$ is driven by the stochastic and self-correlated propulsion force $\eta(t)$:
\begin{subequations}
	\begin{align}
		\zeta\dot x &= f(x) + \eta \ ,
			\label{eqn:LE_dimx}
		\intertext{where the force $f(x)$ is the derivative of an externally imposed potential $U(x)$. In these dynamics, the friction term $\zeta \dot{x}$ is instantaneous and memoryless, while the force $\eta(t)$ is correlated on time-scale $\tau$. Hence the fluctuation-dissipation relation between the friction kernel and force correlation is explicitly violated.\newline\indent
		To describe the fluctuations and dynamics of the propulsion force, we imagine that it is itself driven by a hidden Gaussian white noise $\xi(t)$, such that}
		\tau\dot\eta &= -\eta + \xi(t) \ ,
			\label{eqn:LE_dimeta}
	\end{align}
	\label{eqn:LE_dim}
\end{subequations}%
where $\xi(t)$ is a white noise force with $\left<\xi(t)\right>=0$ and $\left<\xi(t)\xi(t^\prime)\right>=2\zeta T\delta(t-t^\prime)$, with $T$ denoting the temperature in energy units (that is, where $k_{\rm B}=1$). The coefficient $2\zeta T$ controls the intensity of the white noise, and is chosen such that the standard passive Brownian dynamics is recovered when $\tau=0$.

Equation~(\ref{eqn:LE_dimeta}) indicates that the dynamics of $\eta(t)$ is independent of the particle's position (which is also the case for ABPs and RTPs). It also ensures that its correlation function is exponential: $\left<\eta(t)\eta(t^\prime)\right>=(\zeta T/\tau)\exp\left[-|t-t^\prime|/\tau\right]$.
In appendix~\ref{app:GeneralCorrelation}, we show that a more general situation, where the self-correlation of $\eta(t)$ is characterised by several distinct time-scales, can be treated by simply introducing a (possibly infinite) number of hidden white noise variables.

Having introduced the white noise $\xi(t)$ into equation~(\ref{eqn:LE_dimeta}), we can now consider the time evolution of a ``coarse-grained'' phase space density $\rho(x,\eta,t)$ \cite{vanKampen}. This evolution is governed by the Fokker--Planck Equation (FPE):
\begin{align}
	\partial_t \rho = -\frac{1}{\zeta}\partial_x\left[(\eta+f(x))\rho\right] + \frac{1}{\tau}\partial_\eta\left[\eta\rho\right] +\frac{\zeta T}{\tau^2}\partial_\eta^2\left[\rho\right] \ ,
	\label{eqn:FPE_dim}
\end{align}
where the first two terms on the right-hand side are advections and the last term is diffusion in $\eta$-space.

The FPE~(\ref{eqn:FPE_dim}) was used in \cite{Sza14} to derive the stationary distribution of a single OUP in linear and quadratic potentials.  A superficially equivalent description, favoured by a number of other authors (e.g. \cite{Mar++16, Fod++16}), interprets $\eta/\zeta$ as the particle's \textit{swimming velocity}, while we prefer to view $\eta$ as propulsion force, the difference being their time reversal signature \cite{CN1}.  The $(x,\eta)$ phase space will be our main technical tool in this article.

For simplicity, we may choose units which make the spatial coordinate $x$ and the propulsion force $\eta$ dimensionless, so that equations~(\ref{eqn:LE_dim}) become
\begin{subequations}
	\begin{align}
		\dot x &= f(x) + \eta
			\label{eqn:LE_x}\\
		\alpha\dot\eta &= -\eta + \xi(t) \ ,
			\label{eqn:LE_eta}
	\end{align}
	\label{eqn:LE}
\end{subequations}
where $\alpha$ is the dimensionless correlation time, $\left<\eta(t)\right>=0$ and $\left<\eta(t)\eta(t^\prime)\right>=({1}/{\alpha})\exp\left[-|t-t^\prime|/\alpha\right]$.
The FPE corresponding to equation~(\ref{eqn:LE}) is
\begin{align}
	\partial_t \rho = -\partial_x\left[(\eta+f(x))\rho\right] + \frac{1}{\alpha}\partial_\eta\left[\eta\rho\right] +\frac{1}{\alpha^2}\partial_\eta^2\left[\rho\right] \ .
	\label{eqn:FPE}
\end{align}

In this paper we shall frequently consider quadratic potentials with spring constant $k$. The convenient units to recover dimensionless equations~(\ref{eqn:LE}) and~(\ref{eqn:FPE}) are then $\text{length}=\sqrt{T/k}$, $\text{force}=\sqrt{Tk}$ and $\text{time}=\zeta/k$. Thus the dimensionless correlation time $\alpha\equiv{\tau k}/{\zeta}$. For clarity, we shall henceforth work in these units, so all variables are dimensionless and the distance from equilibrium is parameterised by $\alpha$.

From the correlation function for $\eta(t)$, we may infer a characteristic propulsion force $1/\sqrt{\alpha}$, which in turn suggests a characteristic length-scale $\sqrt{\alpha}$ over which the propulsion of a \textit{free} OUP remains correlated. These scales will come in useful for interpreting OUP behaviour.
For instance, at any point which is many correlation lengths away from the nearest potential -- we call this ``deep in the bulk'' -- the spatial derivative in equation~(\ref{eqn:FPE}) disappears, and the dynamics is that of an Ornstein--Uhlenbeck process with steady-state distribution
\begin{align}
	\rho(^\text{deep}_\text{bulk},\eta)\propto\exp\left[-\frac{1}{2}\alpha\eta^2\right] \ .
	\label{eqn:eta_deep_in_bulk}
\end{align}

\section{OUPs and Passive Particles}
\label{sec:general}

In this section, we highlight two cases where the OUP model can be linked to the dynamics of passive particles. First, we discuss a superficial similarity between the OUP and a Brownian particle in a shear flow. Second, we detail how the limit of small correlation times gives rise to passive Brownian dynamics.

\subsection{Analogy with a passive particle in a shear flow}
\label{app:shear}

The OUP model with a quadratic external potential in one spatial dimension bears some similarity, as well as a fundamental difference, with the two-dimensional dynamics of an overdamped Brownian particle which is confined by a harmonic potential and experiences linear shear. For instance, we can imagine a Brownian particle attached by a spring (with spring constant $k$) to the origin and subject to a shear flow with velocity $v_x = \gamma y$, where $\gamma$ is the shear rate. Such a system (if overdamped) is described by a pair of Langevin equations
\begin{subequations}\begin{align}
	\zeta \dot{x} & = - k x + \gamma y + \xi_x(t) \ , \\ \zeta \dot{y} & = - k y + \xi_y(t) \ ,
\end{align}\end{subequations}
where $\xi_x(t)$ and $\xi_y(t)$ are two mutually independent white noises: $\left< \xi_a(t) \right> = 0$ and $\left< \xi_a(t) \xi_b(t^{\prime}) \right> = 2 \zeta T \delta_{ab} \delta(t-t^{\prime})$.  This system exhibits some analogy with our system: in both cases, the system arrives with time to the steady state with two-dimensional Gaussian density distribution \cite{R+Z03}.  Moreover, it is easy to find shear rate $\gamma$ at which elliptic level lines of the two-dimensional density will have the same inclination.

Importantly, neither system comes to equilibrium, and their respective steady states are characterised by loopy currents (which in the case of the passive particle is simply tumbling).

But the analogy, however appealing, stops at this point.  In the shear flow case, the system does not come to equilibrium because it is subject to a non-conservative (non-potential) field of forces (compare to \cite{BrownianVertex_3}).  By contrast, OUPs do not come to equilibrium for a fundamentally different reason, because their memory-less friction and correlated propulsion force violate fluctuation-dissipation.  Mathematically this is reflected in the absence of white noise terms in the first of the two Langevin equations, (\ref{eqn:LE_x}).

\subsection{OUP Diffusion Equation on Long Time-Scales in a Smooth Potential}
\label{app:DiffusionEquation}

The Fokker-Planck equation~(\ref{eqn:FPE}) contains no diffusion in the $x$-direction.  Yet on time-scales much longer than the correlation time $\tau$ (or $\alpha$ in dimensionless units), the correlated noise $\eta(t)$ should act like a white noise.  It is methodically important to understand how diffusion along $x$ arises over long times, but this currently appears to be missing from the OUP literature.

In $d$ dimensions equation~(\ref{eqn:LE_x}) governing the dynamics of the OUP position $\vec x$ reads
\begin{align}
	\dot{\vec x} = \vec f(x) + \vec\eta(t) \ .
\end{align}
We can transform this into an advection equation for the time-dependent density $n(\vec x)$:
\begin{align}
	\partial_t n(\vec x,t) &= -\nabla\cdot\left[\left(\vec f(\vec x)+\vec\eta(t)\right)n(\vec x,t)\right] \ .
\end{align}
For any advection equation, translation in time is equivalent to a translation in space along a characteristic. The solution must therefore obey
\begin{align}
	n(\vec x, t+\delta t) &= n\left(\vec x - \int_t^{t+\delta t}\left(\vec f(\vec x)+\vec \eta(t^\prime)\right)\upd t^\prime, t\right),
	\intertext{provided the time-scale of interest, $\delta t$, is small enough that $\vec f(\vec x)$ does not change substantially due to the particle's motion (this is discussed further below).
	Provided this condition is satisfied, we can Taylor expand to second order}
	\begin{split}
	n(\vec x, t+\delta t) &\simeq n(\vec x,t) + \\
		&- \nabla \cdot \left[ \int_t^{t+\delta t}\left(\vec f(\vec x)+\vec \eta(t^\prime)\right)\upd t^\prime n(\vec x,t)\right] +\\
		&+ \frac{1}{2}\int_t^{t+\delta t}\left(\vec f(\vec x)+\vec \eta(t^\prime)\right)\upd t^\prime \times\\
		&\quad\times\int_t^{t+\delta t}\left(\vec f(\vec x)+\vec \eta(t^{\prime\prime})\right)\upd t^{\prime\prime}\cdot \nabla^2 n(\vec x,t) \ .
		\label{eqn:DE_characteristic_expansion}
	\end{split}
\end{align}

As stated above, we are interested in situations where the \textit{correlation time is small} compared to the time-scale of interest: $\alpha\ll\delta t$. Then $\vec\eta(t)$ samples much of the available state space in time $\delta t$, and we are justified in taking an average over the random variable. The mean $\left<\vec\eta(t)\right>=\vec0$, while the correlator $\left<\vec\eta(t^\prime)\cdot\vec\eta(t^{\prime\prime})\right> = (d/\alpha)\exp\left[-|t^\prime-t^{\prime\prime}|/\alpha\right]$. The average of the first integral of equation~(\ref{eqn:DE_characteristic_expansion}) is therefore easy, while the second integral takes a couple of lines and equals $\alpha\,\delta t+\mathcal O(\alpha^2)$. Ignoring terms proportional to $\delta t^2$, we obtain (in regular units)
\begin{align}
	\begin{split}
	\partial_t n(\vec x, t) &\simeq - \frac{1}{\zeta}\nabla \cdot \left[ \vec f(\vec x) n(\vec x,t)\right] + D \nabla^2 n(\vec x,t) \ ,
	\end{split}
\end{align}
where the diffusivity $D\equiv\frac{2 T d}{\zeta}$. Thus we recover the advection-diffusion equation for passive particles provided the time-scale of interest is larger than the correlation time, and the potential does not vary too quickly.

A more quantitive statement of this last condition will depend on the particular potential under consideration. Generically, we want to ensure that in time $\delta t$ the OUP does not displace a significant distance compared to the smallest length-scale of the external potential. Since we consider the small correlation time regime and $\vec\eta(t)$ is statistically averaged, it is fair to say that the relevant displacement length-scale is the particle's root-mean-square displacement  over time $\delta t$, $\sqrt{\left<\left(\vec x(\delta t)-\vec x(0)\right)^2\right>}$.

\section{Pressure Exerted by a Confined OUP}
\label{sec:pressure}

While it is interesting to know how active OUPs behave like passive particles in a smooth potential landscape over a long time, it is more interesting when there are clear signatures of activity.  Following previous works, we shall now investigate the behaviour of OUPs which exert pressure on relatively sharp potentials.  Nomenclature-wise, we designate regions of constant potential as the ``bulk'', and static potential barriers as ``walls''. In many cases, these walls grow to infinite height, confining the OUP to a finite volume.

Consider a steady-state in one spatial dimension. The mechanical pressure exerted by OUPs can be defined as simply the product of the spatial density at each point on the wall, $n(x)$, with the force the wall exerts on each particle at that point, $-f(x)$, integrated over all points:
\begin{align}
	P = -\int_\text{foot of wall}^\text{top of wall} n(x)f(x) \upd x \ .
		\label{eqn:P}
\end{align}

In higher dimensions, it is natural to generalize formula (\ref{eqn:P}) by performing the integration over a path $\mathcal C$ connecting the top and bottom of the wall:
\begin{align}
    P=-\int_{\mathcal C}n(\vec x)\vec f(\vec x)\cdot\upd\vec\ell \ .
        \label{eq:Pmultidimensional}
\end{align}
In the special case that the ``density of force'' $n(\vec x)\vec f(\vec x)$ can be expressed as the gradient of a scalar field, this integral is independent of the path ${\mathcal C}$.  It is instructive to see that this is always the case in any equilibrium system, as we show below in section~\ref{sec:PressureInEquilibrium}. In section~\ref{sec:moments1d} we shall show that it is also true for OUPs in one dimension, and hence formula~(\ref{eq:Pmultidimensional}) allows us to determine their pressure.  In other cases, as we shall show in section~\ref{sec:momentsdd}, the density of force may not be a potential field, and formula~(\ref{eq:Pmultidimensional}) is not sufficient.
In such cases, coarse graining the system over time scales larger than $\tau$ and over distances larger than $\eta \tau / \zeta$ (with $\eta$ properly averaged), and properly averaging formula~(\ref{eq:Pmultidimensional}) may warrant additional investigation (see also \cite{MURA, SFG17}).

\subsection{Pressure in an Equilibrium System}
\label{sec:PressureInEquilibrium}

To fully appreciate the peculiarities of active OUPs, it is useful to understand how formula~(\ref{eq:Pmultidimensional}) works in the familiar equilibrium scenario.  In the presence of confining potential $u(\vec x)$, the equilibrium density $n(\vec x)$ balances so as to maintain constant total chemical potential, $\mu \left(n (\vec x) \right) + u(\vec x) = \mathrm{constant}$ (see, e.g., \cite[Sec. 25]{LL_volV}).  Taking the gradient of this equation and remembering that ${\vec \nabla} u = - {\vec f}(\vec x)$, we arrive at $\frac{\partial \mu}{\partial n} {\vec \nabla} n = {\vec f}(\vec x)$.  Multiplying now both sides by $n(\vec x)$ and using the thermodynamic relation $n \frac{\partial \mu}{\partial n} = \frac{\partial p}{\partial n}$, we finally discover that
\begin{align}
    n(\vec x) {\vec f}(\vec x) = {\vec \nabla} p \ .
\end{align}
Thus, the vector field $n(\vec x) {\vec f}(\vec x)$ is \textit{conservative}, and the local pressure $p$ plays the role of its potential.  If an equilibrium system is confined by walls, then the material of these walls is subject to the potential force field exerted by our system, and this material itself can be in equilibrium.


\subsection{OUP Pressure in One Dimension}
\label{sec:moments1d}

We now seek the pressure exerted by OUPs in one dimension, and start by finding the \emph{moments} of the $\eta$ distribution,
defining the $m$th moment of $\eta$ {at each point in space}, $\left<\eta^m\right>(x)$, as:
\begin{align}
	\left<\eta^m\right>(x)n(x) \equiv \int_{-\infty}^{+\infty}\eta^m\rho(x,\eta)\upd\eta \ .
\end{align}
Using the steady-state FPE~(\ref{eqn:FPE}) these moments are found to obey the recurrence differential equation
\begin{align}
	\begin{split}
		0 =& -\partial_x\left[\left(\left<\eta^{m+1}\right>+\left<\eta^{m}\right>f\right)n\right] - \frac{1}{\alpha}m\left<\eta^{m}\right>n+\\
		&\qquad +\frac{1}{\alpha^2}m(m-1)\left<\eta^{m-2}\right>n \ ,
	\end{split}
	\label{eqn:GeneralMoments1d}
\end{align}
where $\alpha$ is the dimensionless correlation time-scale as before, and we've omitted all $x$-dependence. For a confined system in the steady state, the first moment is
\begin{align}
	\left<\eta\right>(x)=-f(x) \ ,
	\label{eqn:FPE_m0}
\end{align}
meaning that for any $x$, the average propulsion force exerted by a single particle is balanced by the potential force at that point. The second moment obeys
\begin{align}
	0 = \partial_x\left[\left(\left<\eta^2\right>(x)-f(x)^2\right)n(x)\right] - \frac{1}{\alpha}f(x)n(x) \ .
	\label{eqn:FPE_m1}
\end{align}

Whereas equation~(\ref{eqn:FPE_m0}) represents the balance of forces on a \emph{single} particle, equation~(\ref{eqn:FPE_m1}) can be interpreted as the balance between the propulsion force of an \emph{ensemble} of particles, and the stresses they impart to the surrounding medium. This will be shown to generalise to higher dimensions in equation~(\ref{eqn:FPE_m1_ind}), and a more elaborate ``hydrodynamic'' discussion can be found in \cite{SFG17}. As was shown in \cite{KML14}, the moments of the FPE are independent of the precise dynamics of the active force, and so essentially the same equations were derived in \cite{Sol++15,S+J16} for the ABP model.

To find the total pressure on a \textit{confining} (infinitely high) wall, we substitute equation~(\ref{eqn:FPE_m1}) into the definition of mechanical pressure in equation~(\ref{eqn:P}), and integrate over $x$ from somewhere in the bulk (where $f=0$) to $x=\infty$. Then
\begin{align}
	P = \alpha \left<\eta^2\right>(x)n(x) |^\text{bulk} \ ,
	\label{eqn:PBC}
\end{align}
where the right-hand side is evaluated in the bulk. This result, reminiscent of Bernoulli's Principle and closely analogous to one derived in \cite{S+J16} for ABPs, relates the pressure exerted on a wall to bulk properties -- an equation of state for OUPs in 1D. Crucially, the right-hand side of equation~(\ref{eqn:PBC}) is independent of precisely where in the bulk it is evaluated: this can be proved by setting $f=0$ in equation~(\ref{eqn:FPE_m1}) to find
\begin{align}
	\left<\eta^2\right>(x)n(x) = \mathrm{constant}	\qquad\text{(in bulk)} \ .
	\label{eqn:BC}
\end{align}
This is not a trivial statement, since active particles' density is strongly affected in the vicinity of walls even when no force acts. But the prediction is well supported numerically, as discussed below in section \ref{sec:CAR_DL}.

Note that we consider point-like like OUPs which experience forces but no torques from the external potential. Active particles which \textit{do} feel torques generically do not obey an equation of state, as was shown in \cite{Sol++15,J+B16} for ABPs and RTPs.  However, in this paper we shall stick with the simpler case of point-like particles.

\subsection{Large Bulk vs. Small Bulk}
\label{sec:moments1d_bulk}

Pressure in 1D is fully determined by bulk properties -- and in the limit of a large bulk, it is in fact determined by the ideal gas law for passive particles (see also \cite{CN1,S+J16}).  This follows straightforwardly from formula~(\ref{eqn:PBC}): deep in the bulk, far from boundaries, $n(x)|^\text{bulk} = n$; and from equation~(\ref{eqn:eta_deep_in_bulk}), $\left<\eta^2\right>(x) = 1/\alpha$. Together these yield $P=n$ (or the familiar $P = nT$ in dimensionful units).

When the large-bulk limit is \emph{not} satisfied, and walls do influence each other, the pressure deviates from the passive thermal value according to equation~(\ref{eqn:PBC}). Once again, the central thrust of this paper is to investigate how the proximity of walls influences the pressure, and our findings will allow us to construct examples of arrangements of potentials which experience net forces from the surrounding OUPs.

In order to develop the necessary intuition for how interaction with external potentials affects the state of an OUP, and in particular how it affects the constant defined in equation~(\ref{eqn:BC}), we shall consider in section~\ref{sec:GO} a simplified model of an OUP interacting with a single wall in 1D. First, however, it is convenient to briefly discuss the pressure exerted by OUPs in higher dimensions.

\subsection{Pressure in Higher Dimensions}
\label{sec:momentsdd}

We shall show in this section that, in contrast to the 1D case just discussed in section~\ref{sec:moments1d}, pressure in higher dimensions is generally \emph{not} a function of bulk properties. 

We start by writing the steady-state FPE in $d$ dimensions as
\begin{align}
	0 &= -\partial_i\left[\left(\eta_i+f_i(\vec x)\right)\rho\right] + \frac{1}{\alpha}\nabla_i\left[\eta_i\rho\right] + \frac{1}{\alpha^2}\nabla_i\nabla_i\left[\rho\right] \ ,
		\label{eqn:FPE_ind}
\end{align}
where Roman indices denote vector components, repeated indices are summed over, $\partial$ denotes a spatial derivative, and $\nabla$ denotes a derivative with respect to propulsion force. Following the same procedure as for 1D, we find $\left<\eta_i\right>(\vec x)=-f_i(\vec x)$ (analogous to equation~(\ref{eqn:FPE_m0})), and
\begin{align}
	0 &= -\partial_i\left[\left(\left<\eta_i\eta_j\right>-f_i f_j\right)n\right] + \frac{1}{\alpha}f_j n
		\label{eqn:FPE_m1_ind}
\end{align}
(analogous to equation~(\ref{eqn:FPE_m1})).

Intriguingly, the first term of equation~(\ref{eqn:FPE_m1_ind}) has a non-zero curl: thus it does not represent a potential vector field, and hence it renders equation~(\ref{eq:Pmultidimensional}) insufficient to define pressure in general. There are two major exceptions to this rule, when the curl of the first term in equation~(\ref{eqn:FPE_m1_ind}) happens to be zero.  First, in geometries where $\vec f(\vec x)$ varies only along one Cartesian dimension (which trivially includes the case of $d=1$), the offending term \textit{can} be expressed as the gradient of a potential field. The pressure is therefore well-defined -- and moreover obeys an equation of state. The second case where pressure is a meaningful quantity is in radially-symmetric geometries. This will be explored further in section~\ref{sec:polar} for $d=2$, where we show that there is no equation of state.

Note that we started with a \textit{mechanical} definition of pressure -- i.e. we consider the potential force required to support a given distribution of particles. Others have examined the correspondence between this formulation and statistical mechanical definitions: see for instance the works \cite{Sol++15,WWG15} on ABPs/RTPs, or \cite{MURA, SFG17} and especially \cite{MMM16} for the OUP model close to equilibrium (we shall make some further comments on the approach of \cite{MMM16} in section~\ref{sec:polar}).

\section{Zero-Diffusion Approximation in 1D}
\label{sec:GO}

Returning to one spatial dimension, we now introduce a simplified model in order to explore the influence of potentials on OUP behaviour. The qualitive insights thus obtained will help us interpret phenomena observed in simulations of the full model.

Consider an OUP with propulsion force $\eta$ which brings it to a confining wall, where the potential grows rapidly. From equation~(\ref{eqn:LE_x}), we see that it will penetrate into the wall region only until its $\eta$ is balanced by the potential force -- so if the wall in question is very steep, the particle's $\eta$ must be very large to substantially penetrate. In such circumstances, which in dimensionless units correspond to $\alpha\gg1$, the dynamics of $\eta$ within the wall region are dominated by drift rather than diffusion, and so we may simplify the OUP problem by neglecting the final diffusive term in the FPE~(\ref{eqn:FPE}).

The dynamics of a system in one spatial dimension are characterised by the flow field in the $(x,\eta)$ plane. The components of the current are given by
\begin{align}
	\vec J(x,\eta) =
		\begin{pmatrix}
			\eta+f(x) \\
			-\frac{1}{\alpha}\eta
		\end{pmatrix}
	\rho(x,\eta).
		\label{eqn:GO_J}
\end{align}

In some cases, useful for illustration, we can solve the trajectories exactly. Defining the velocity field $\vec v(x,\eta)=\vec J/\rho$, particle trajectories are described by
\begin{align}
	\frac{\upd \eta}{\upd x} = \frac{v_\eta}{v_x} = -\frac{1}{\alpha}\frac{\eta}{\eta+f(x)},
\end{align}

\begin{figure}[ht]
	\centering
	\includegraphics[width=\wid\textwidth]{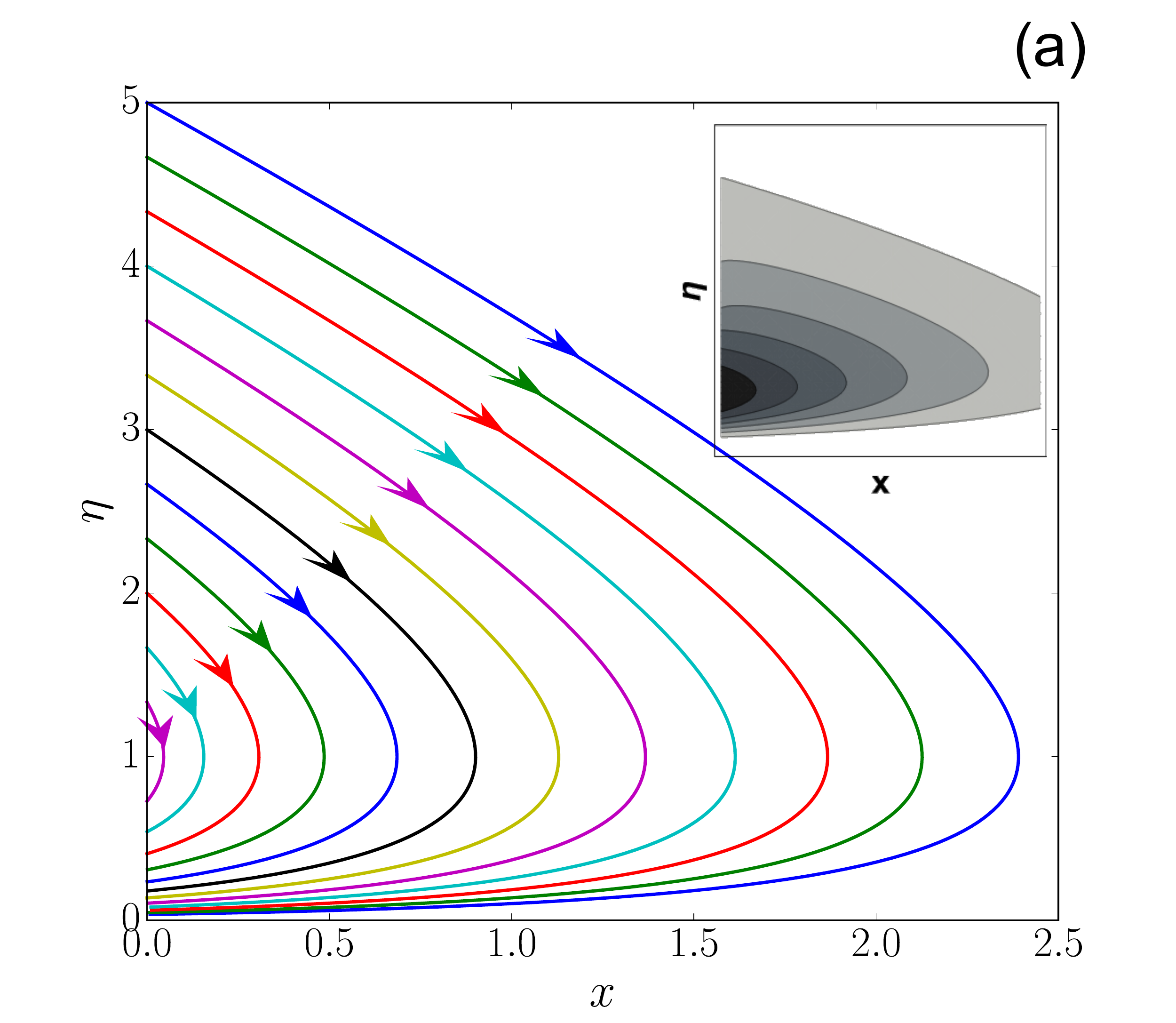}
	\includegraphics[width=\wid\textwidth]{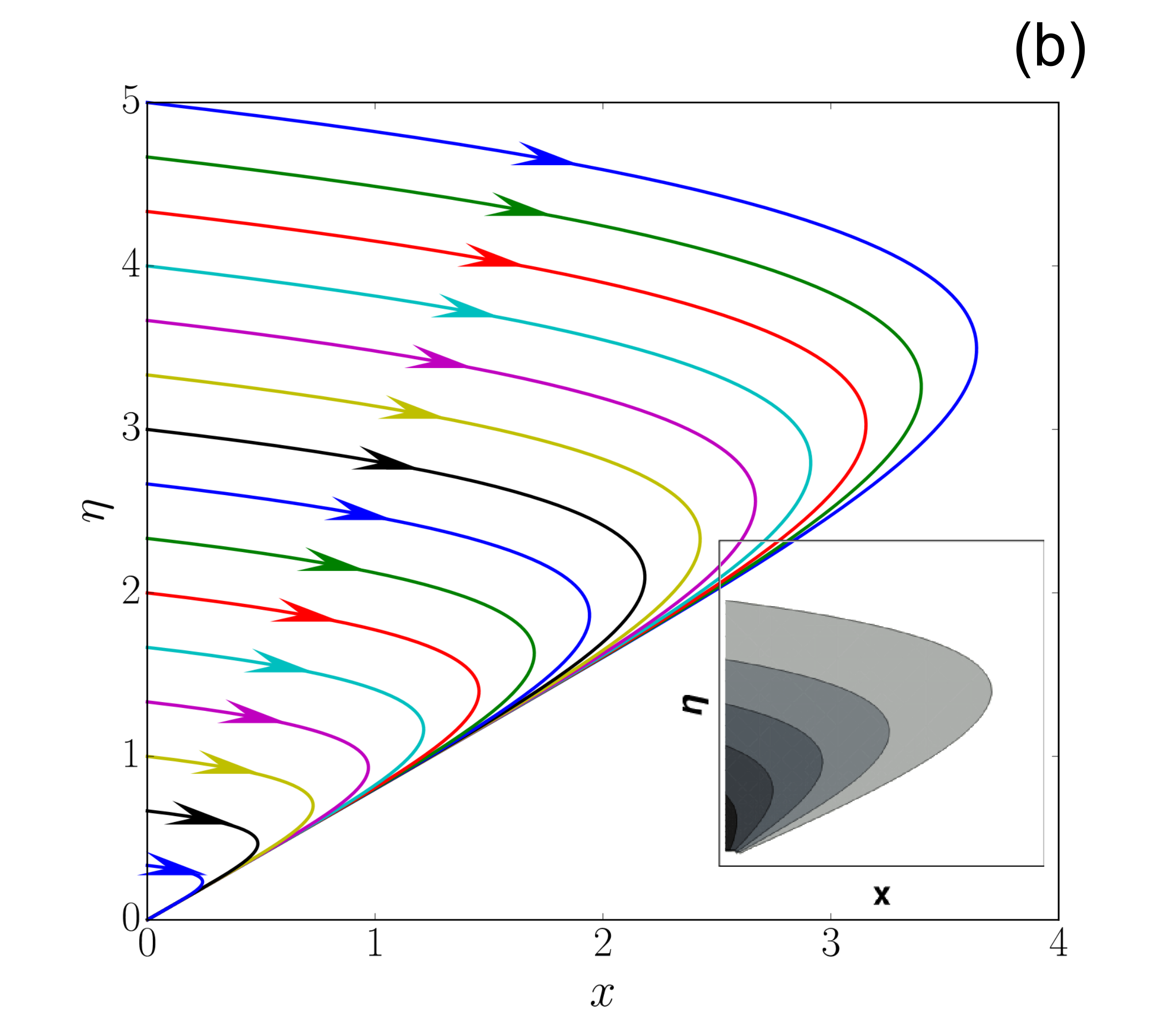}
	\caption{Trajectories in $(x,\eta)$ for the diffusion-free approximation in the the wall region. Each line corresponds to a different $\eta_0$. Insets show the resulting phase space density (darker shading means higher density). \textbf{(a)} linear potential $U=x$. \textbf{(b)} quadratic potential $U=\frac{1}{2}x^2$.}
	\label{fig:GO_traj}
\end{figure}

Let us choose coordinates such that the confining potential spans the region $x>0$.
If the potential is linear, we pick units such that $f(x\geq0)=1$ and $f(x<0)=0$, and the dimensionless correlation time $\alpha\equiv \frac{\tau f_0^2}{\zeta T}$ (where $f_0$ is the slope of the potential). An OUP which enters the wall region at $(x=0,\eta_0)$ will follow the trajectory
\begin{align}
	x(\eta) &= \alpha\left(\eta_0-\eta+\ln\left[\frac{\eta}{\eta_0}\right]\right)	\quad\text{linear potential} \ .
		\label{eqn:GO_traj_C}
	\intertext{If the potential is instead quadratic, and $\alpha\equiv \frac{\tau k}{\zeta}$ as defined in section~\ref{sec:model}, the trajectory is}
	x(\eta) &= \frac{\alpha}{\alpha-1}\eta\left(1-\left(\frac{\eta}{\eta_0}\right)^\alpha\right)	\quad\text{quadratic potential}
		\label{eqn:GO_traj_L}
\end{align}
(remember we must have $\alpha\gg 1$ for this approximation to apply). The trajectories~(\ref{eqn:GO_traj_C}) and~(\ref{eqn:GO_traj_L}) are plotted parametrically in Fig.~\ref{fig:GO_traj}.

The insets in Fig.~\ref{fig:GO_traj} show the phase-space density in the wall region, which was found by invoking continuity of current along a trajectory. To perform this calculation, we assumed that OUPs entered the wall region with $\eta$ drawn from the deep-bulk Gaussian distribution (\ref{eqn:eta_deep_in_bulk}) -- a somewhat inconsistent choice in the high-$\alpha$ regime where there is significant distortion of the $\eta$ distribution close to the wall. Nevertheless, there is qualitative agreement with the density found by numerical simulations, to be discussed in the following section.

However, the main physical insight derived from this diffusion-free model is evident from the trajectories in Fig.~\ref{fig:GO_traj} and concerns the effect of walls on a particle's $\eta$: OUPs ejected from the wall region tend to have ``depleted'' $\left<\eta^2\right>$, meaning they move slowly and contribute to a density peak in the vicinity of the wall.

\section{An OUP Confined in a Finite Bulk}
\label{sec:CAR_DL}

The next two sections present the results of single-OUP simulations in one spatial dimension (though as discussed in section~\ref{sec:momentsdd}, the results apply equally well to quasi-one-dimensional channels, where $U(\vec x)=U(x)$). The simulation consists of integrating the dimensionless stochastic equations~(\ref{eqn:LE}) until convergence to the steady state.

In section~\ref{sec:moments1d}, we found that in 1D there is both a sensible definition of pressure and an equation of state~(\ref{eqn:PBC}). But as we shall see, there are still a number of interesting consequences of the OUPs' activity.

\begin{figure}[ht]
	\centering
	\includegraphics[width=\wid\textwidth]{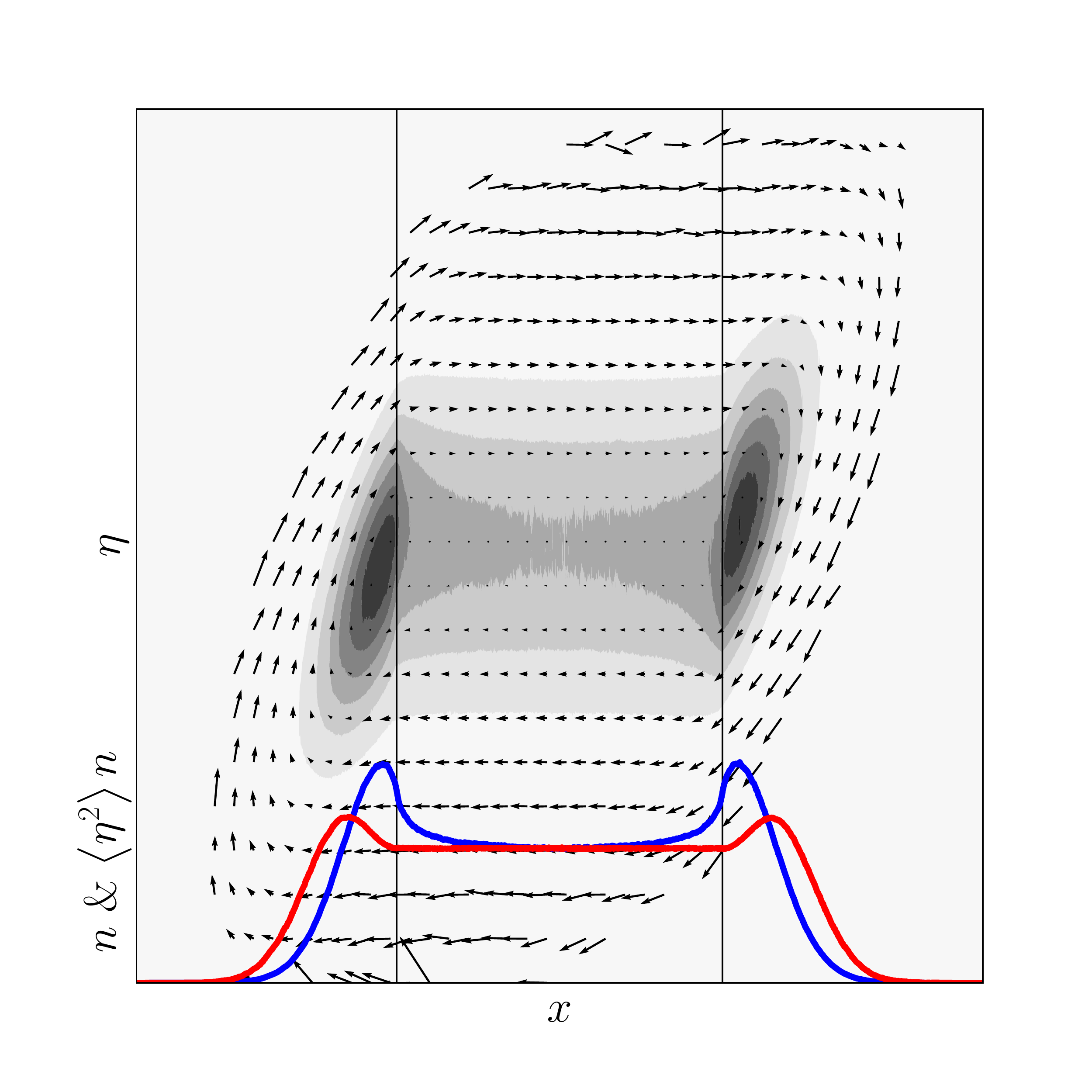}
	\caption{Contours: density $\rho(x,\eta)$ for an OUP confined by quadratic walls with a finite bulk of zero force in between (this region is demarcated by vertical lines). Vectors: velocity field. Curves: projected spatial density $n(x)$, which varies in the bulk, and the quantity $\left<\eta^2\right>(x)n(x)$ of equation~(\ref{eqn:BC}), which is indeed constant in the bulk.}
	\label{fig:PDF_CAR_DL_FB}
	\label{fig:BC_CAR_DL_FB}
\end{figure}

It has been shown previously that when an OUP is confined in 1D a quadratic potential, its steady-state density in $(x,\eta)$-space is a bivariate Gaussian \cite{Sza14,CN1}. Let us now consider a potential landscape where there is a finite bulk region (where the external force is zero) between two piecewise-quadratic confining walls. The simulated density and velocity field is shown in Fig.~\ref{fig:PDF_CAR_DL_FB}.

Equation~(\ref{eqn:BC}) predicts that the product $\left<\eta^2\right>(x)n(x)$ is a constant in the bulk, regardless of where it is evaluated. This is verified by the red curve in figure~\ref{fig:BC_CAR_DL_FB}, which remains flat all the way up to the wall even though the density $n(x)$ (blue curve) is changing.

Since the quantitiy $\left<\eta^2\right>(x)$ is related to the pressure through equation~(\ref{eqn:PBC}), it is interesting to know how it might be influenced by the size of the bulk region between the walls. The main finding of the previous section~\ref{sec:GO} was that $\left<\eta^2\right>$ diminishes in the vicinity of walls. Thus, we should expect that closely-spaced walls will diminish the OUP pressure. Furthermore, it is reasonable to suppose that this effect will be at its strongest when the width of the bulk is smaller than the free-particle correlation length, $L\lesssim \sqrt\alpha$, such that the OUP has insufficient time to recover its depleted propulaion force between interactions with walls. (It bears pointing out that such finite-size effects are routinely observed in systems of \textit{interacting passive} particles. But whereas these correlations are induced by long-range forces, for OUPs they are a result of the particle's memory.)

\section{A Penetrable Interior Wall}
\label{sec:CAR_ML}

The far-reaching influence of potentials on the statistics of OUP propulsion suggests that, even in the steady state, an arrangement of identical walls separated by bulks of unequal size will result in net forces -- something which is forbidden for passive particles in equilibrium (in the absence of long-range interactions).

Active pressure imbalances have been investigated before, for instance in \cite{Sol++15}, where the alignments of elliptical ABPs were affected differently by potentials on either side of a partition. This was shown to destroy the equation of state, and resulted in translation of the partition.

However, in the situations to be investigated here, the physics is fundamentally different. We consider point-like particles which experience no torques, so the equation of state~(\ref{eqn:PBC}) is preserved. The net forces arise not from OUPs' specific interactions with walls, therefore, but from {the dissipating memory of spatial correlations that these interactions induce}. Put simply, we observe the influence of the \emph{gaps} between the walls, rather than the walls themselves.

\begin{figure}[ht]
	\centering
	\includegraphics[width=\wid\textwidth]{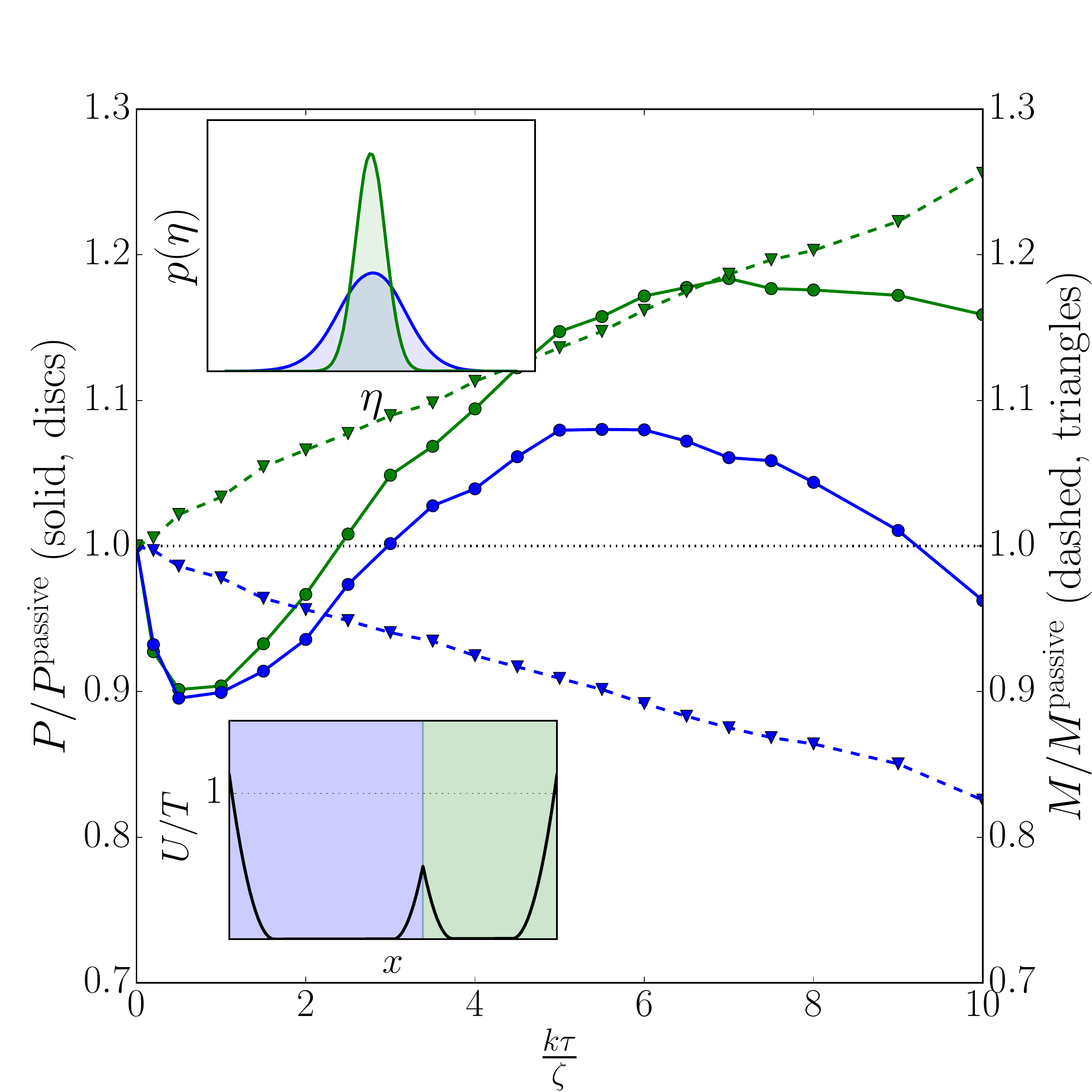}
	\caption{
	{Main figure:} Pressure (circles, solid) and total probability (triangles, dashed) on either side of the penetrable inner wall, plotted against $\alpha\equiv k\tau/\zeta$. The small bulk experiences larger pressure and accumulates more particles.
	{Upper inset:} $\eta$ distribution in the bulk of the two regions. The small bulk has a narrower distribution.
	{Lower inset:} A sketch of the potential.}
	\label{fig:PA_CAR_ML}
	\label{fig:M_CAR_ML}
\end{figure}

More concretely, we performed simulations of an OUP confined to a volume featuring a small, penetrable interior wall placed asymmetrically between the confining walls (see illustration in the lower inset of Fig.~\ref{fig:PA_CAR_ML}). The solid lines in Fig.~\ref{fig:PA_CAR_ML} plot the pressure on either side of the interior wall. For $\alpha>0$, a pressure difference develops in the direction of the larger bulk, such that if the interior wall was mobile, stable mechanical equilibrium would be established only in the case of spatial symmetry (and since the wall is penetrable, the total OUP mass on either side would be equal, in contrast with the ABP simulations in \cite{Sol++15}).

That the pressure is higher on the side of the small bulk seems to contradict the conclusions of section~\ref{sec:CAR_DL}, where it was argued that the OUP pressure on confining walls \textit{grows} with the size of the bulk.
The solution of this conundrum is the penetrability of the interior wall, which causes the smaller bulk to act like a ``particle trap'' (see also Ref. \cite{CN1}). In section~\ref{sec:GO}, we found that walls do indeed sap OUPs' propulsion forces, which diminishes their penetration; but in this context the outcome is that particles in the small bulk are too feeble to overcome the force barrier and escape. This is not the case for the neighbouring large bulk, which shoots high-propulsion particles over the interior wall and into the small bulk.
Hence, mass accumulates disproportionately in the smaller bulk, and this outweighs the effect of diminished propulsion and penetration.

Also of interest is the non-monotonicity of both pressures in Fig.~\ref{fig:PA_CAR_ML}, which can be understood as a competition between decreasing penetration and force-controlled accumulation. We can roughly divide the domain into three regions: $\alpha \lesssim 1$, where the pressure on both sides of the interior wall is lower than the thermal value and falls with $\alpha$; the region $1\lesssim\alpha\lesssim5$ where the pressure grows with $\alpha$; and the region $5\lesssim \alpha$ where the pressure once again falls with $\alpha$.
In the first region, we are seeing the effects of diminished penetration, while in the second, accumulation around the interior wall is the dominant factor (since the wall is penetrable, the accumulation on one side of the wall reinforces that on the other side). In the third region, $\alpha$ is sufficiently high (and the penetration sufficiently low) that the two wells become isolated, and the accumulation on either side no longer reinforces. Then we are approaching to the situation from section~\ref{sec:CAR_DL} (times two) where the pressure on two confining walls was equal and fell with $\alpha$.

Taking all four walls into account (see Fig.~\ref{fig:PAall_CAR_ML} in appendix~\ref{app:CAR_ML_netforce}), the system experiences a net force in the direction of the larger bulk, provided the potential is permeable to the medium. Since only the OUPs are free to move, the entire system will then translate in this direction.
(This contrasts superficially with the ABP setup considered in \cite{Sol++15}, for which the system translates only until mechanical equilibrium is established.)

Further consequences of these phenomena are explored in \cite{CN1}, which describes a repulsive interaction between walls in a Casimir-type setup.

\section{Geometrical Curvature}
\label{sec:CAR_UL}

We have seen in the previous section that an asymmetrical placement of walls may result in OUPs exerting unbalanced forces even in 1D. The diversity of possible asymmetries broadens in higher dimensions to include spatial curvature. The nontrivial interactions of ABPs and RTPs with curved hard walls has previously been investigated theoretically and in simulations \cite{FBH14,FBH15,S+L15,Nik++16}, as well as in experiments \cite{Gal++07,PhysRevE.89.032720}.

\begin{figure}[ht]
	\centering
	\includegraphics[width=\wid\textwidth]{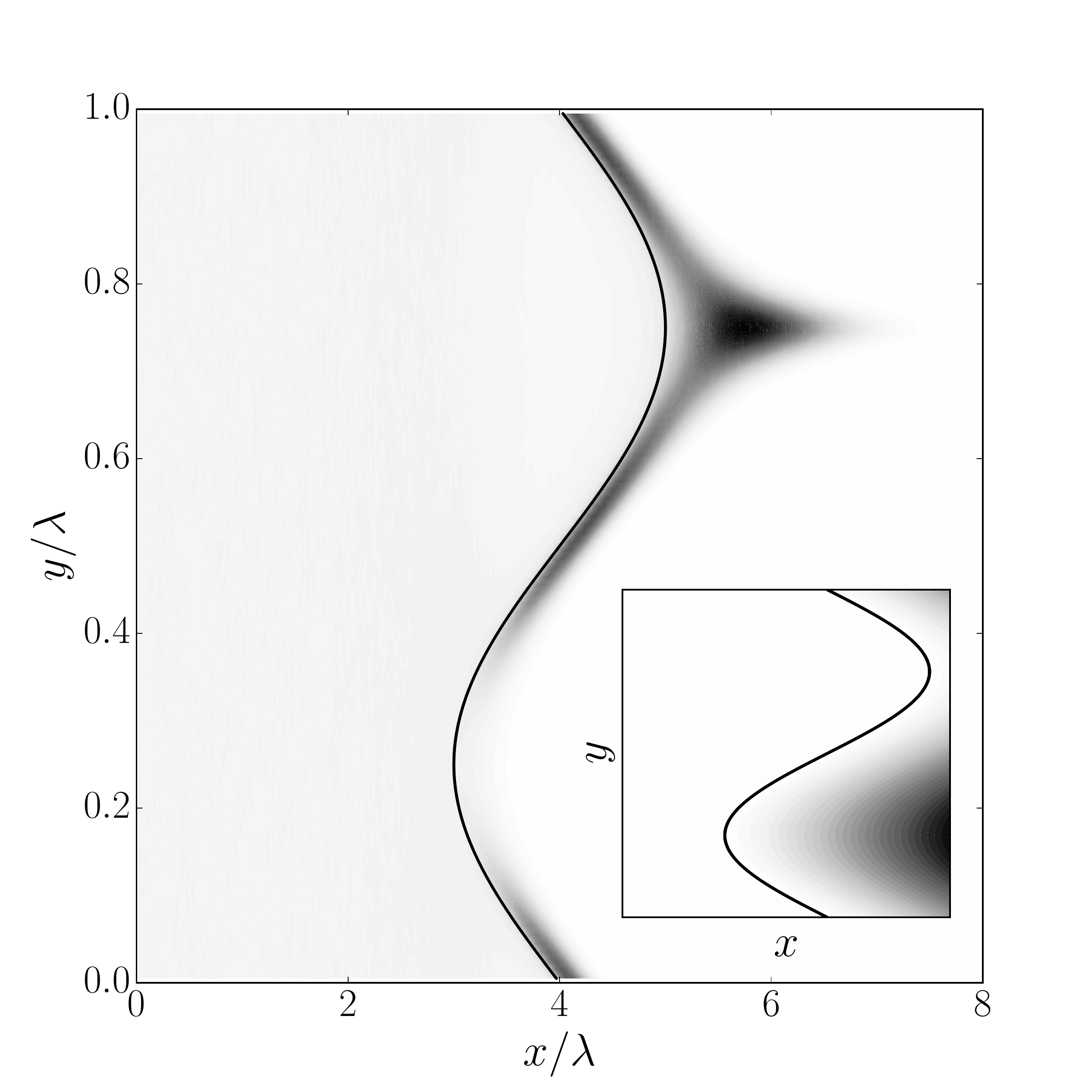}
	\caption{
	{Main figure:} Projected density $n(x,y)$, with the bulk region on the left and darker colour corresponding to higher density.
	{Inset:} Potential (periodic along $y$, with period $\lambda$), where darker colour corresponds to higher potential.}
	\label{fig:Q_CAR_UL}
\end{figure}

First we shall consider a potential analogous to the one in \cite{Nik++16}, where a bulk region is flanked by sinusoidal equipotentials (sketched in the inset of Fig.~\ref{fig:Q_CAR_UL}),
\begin{align}
	U(x,y) =
		\begin{cases}
			&\frac{1}{2}\left(x-R+A\sin\left(2\pi\frac{y}{\lambda}\right)\right)^2 \\ &\qquad\qquad\text{for } x>+R-A\sin\left(2\pi\frac{y}{\lambda}\right) \\
			&\frac{1}{2}\left(x+R+A\sin\left(2\pi\frac{y}{\lambda}\right)\right)^2 \\ &\qquad\qquad\text{for } x<-R-A\sin\left(2\pi\frac{y}{\lambda}\right) \\
			&0 \qquad\quad\;\;\text{elsewhere}\\
		\end{cases}
	\label{eqn:U_CAR_UL}
\end{align}
where $R$ is the mean position of the wall, $A$ is the amplitude of the sinusoid, and $\lambda$ is the period. For now we shall consider wall separations $R$ so large that the bulk is essentially infinite -- that is, $R\gg\sqrt\alpha$.

\begin{figure}[ht]
	\centering
	\includegraphics[width=\wid\textwidth]{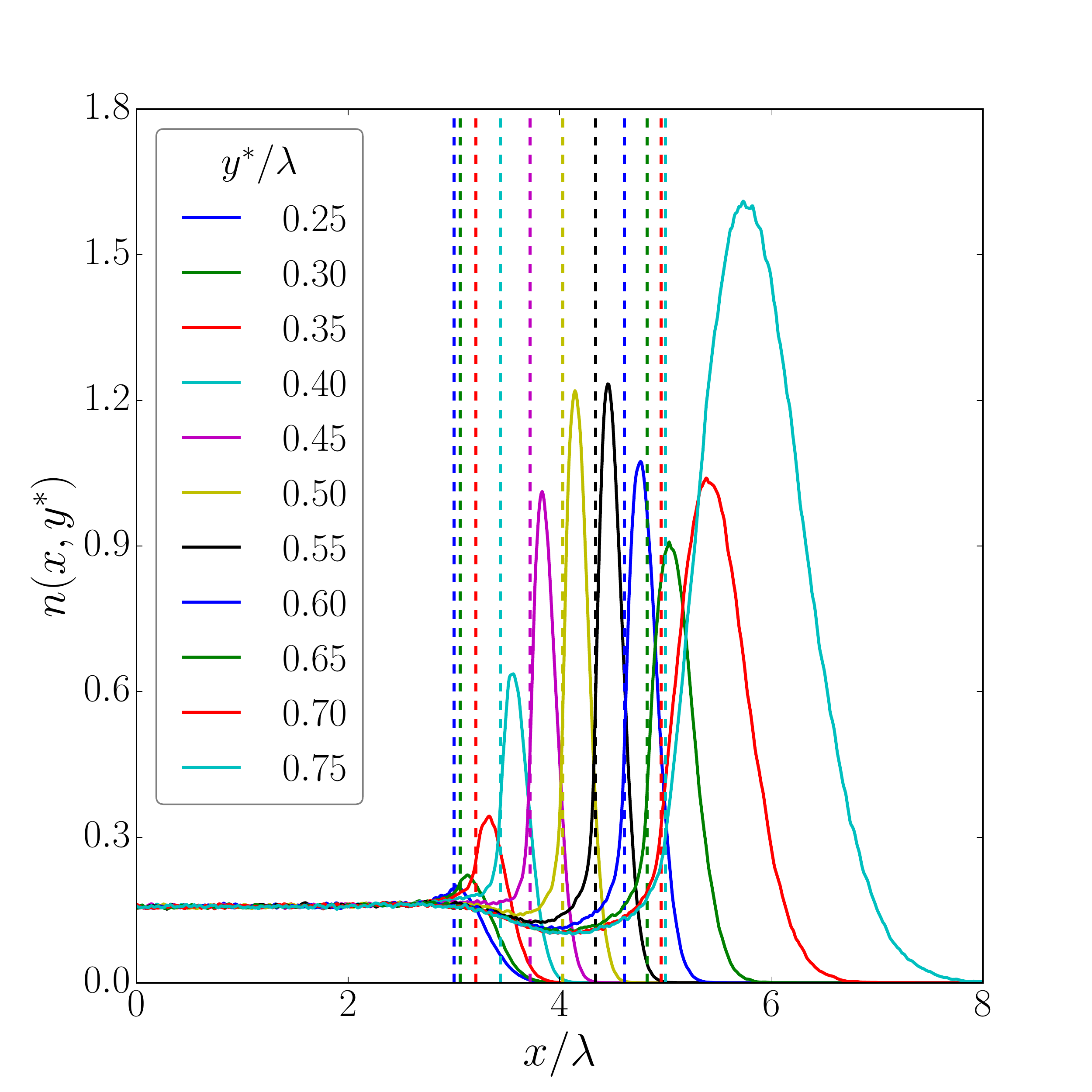}
	\caption{Slices of density at fixed coordinate $y$, from the convex apex ($y=0.5\lambda$) to the concave apex ($y=0.75\lambda$), where $\lambda$ is the period of the wall. Peaks move progressively to the right with increasing $y$, as one would expect. The $x$-position of the wall is indicated by a dashed line of matching colour.}
	\label{fig:Qslice_CAR_UL}
\end{figure}

The simulated density $n(x,y)$ is shown in Fig.~\ref{fig:Q_CAR_UL}.
We can immediately see that force-controlled accumulation along boundaries is still a dominant feature. Like ABPs and RTPs confined by hard walls \cite[supplementary material]{Nik++16}, OUPs concentrate at the concave apex ($y=0.75\lambda$), and are depleted in the surrounding bulk (see also Fig.~\ref{fig:Qslice_CAR_UL}).

These observations are intuitively reasonable, and are most easily explained in the large-persistence limit, where the time for a swimmer to travel the length of the wall boundary is less than or comparable to the correlation time \cite{FBH14,FBH15}.
In the bulk of the concave region ($y>0.5\lambda$ and $3\lambda< x< 4\lambda$ in Fig.~\ref{fig:Q_CAR_UL}) particles are likely to be ``captured'' by the surrounding walls, depleting the density in this region.
Moreover, particles which approach the wall boundary with $\vec \eta$ pointing towards the concave apex (relative to the local wall normal) will be channeled towards that apex until their $\vec\eta$ is matched by the opposing potential force $\vec f$ -- which happens when $\vec\eta$ is aligned with the wall normal. Having reached a mechanical equilibrium, these particles linger and contribute to the pressure until their $\vec\eta$ changes substantially enough to take them off the wall. If on the other hand, particles approach the wall boundary with $\vec\eta$ pointing \textit{away} from the concave apex, their $\vec\eta$ will never be balanced by $\vec f$ and they eventually shoot off the wall into the bulk.

Since OUPs exert a propulsion force of variable magnitude on the {soft} walls, the density at a given point of the wall cannot be simply related to its local curvature, even in the high-persistence limit (as was the case in \cite{FBH15}). An illustration of this fact is visible in Fig.~\ref{fig:Qslice_CAR_UL}, where both the peak density and the density for a given distance into the wall region is non-monotonic along the wall.

\begin{figure}[ht]
	\centering
	\includegraphics[width=\wid\textwidth]{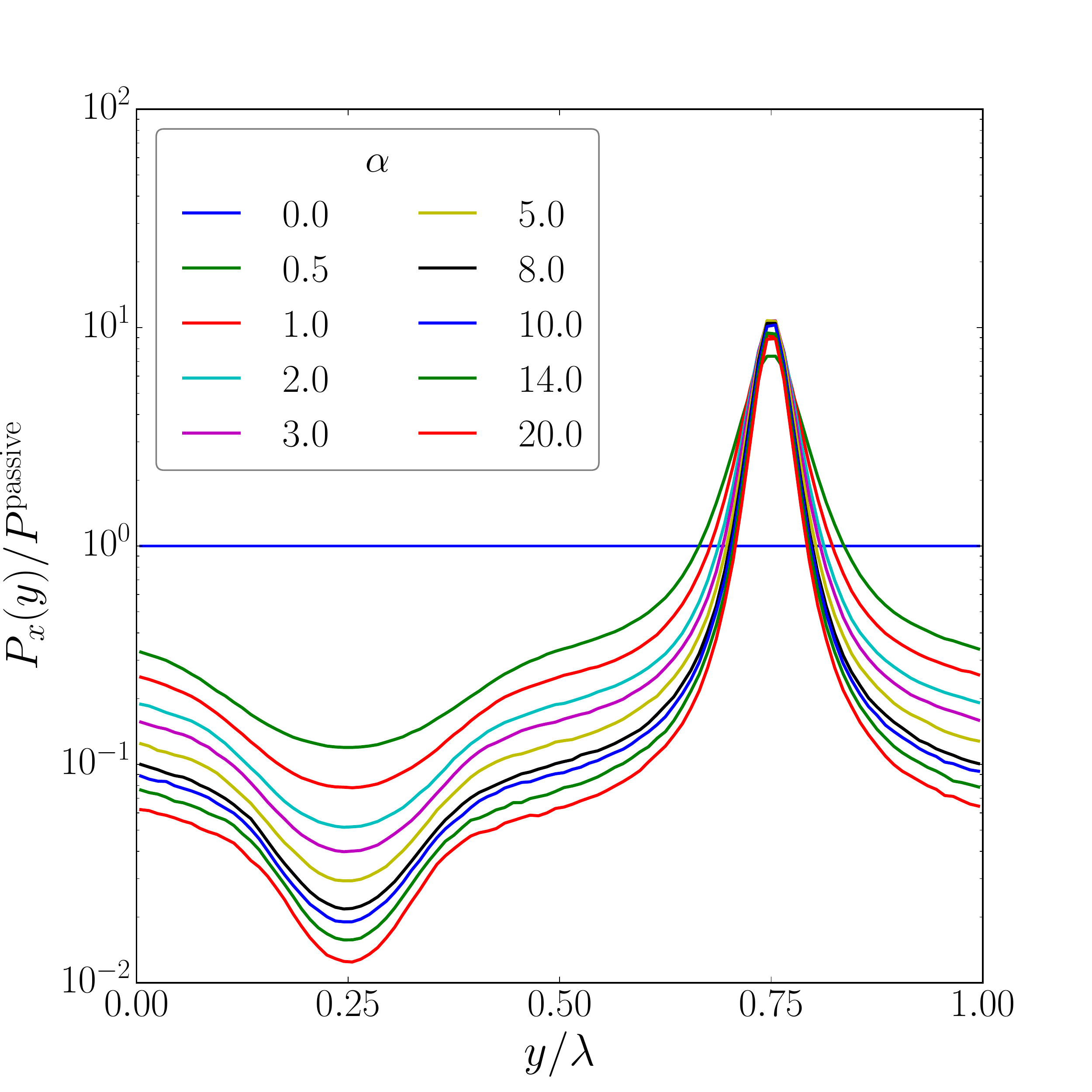}
	\caption{Stress in the $x$-direction as a function of $y$ (defined in equation~(\ref{eqn:Py})). The $\alpha=0$ pressure is constant, and lines progress monotonically with increasing $\alpha$. Note logarithmic ordinate.}
	\label{fig:Py_CAR_UL}
\end{figure}

In section~\ref{sec:momentsdd} we showed that in multi-dimensional geometries such as this one, the OUP pressure calculated according to equation~(\ref{eq:Pmultidimensional}) is path-dependent, and therefore the terms ``pressure'' (or ``stress'') must be accompanied by the understanding that we have invoked some provisional definition.
Following \cite{Nik++16}, we consider the following active stress on a soft wall in the $x$-direction:
\begin{align}
	P_x(y) = -\int_\text{bottom of wall}^{\infty} f(x,y)n(x,y)\upd x.
	\label{eqn:Py}
\end{align}
Fig.~\ref{fig:Py_CAR_UL} shows that between the two apices of the wall this stress is monotonic, and broadly resembles the analogous results for ABPs confined by hard walls \cite{Nik++16}.

It was furthermore shown in \cite{Nik++16} that for ABPs, the stress $P_x(y)$ integrated over a period of any periodic wall is equal to the (uniform) pressure experienced by a flat wall. The same is true for OUPs, as shown explicitly in appendix~\ref{app:pressure_periodic}. Therefore, on spatial scales larger than the periodicity $\lambda$ an equation of state is recovered once more.


\section{Radially Symmetric Geometry}
\label{sec:polar}

Geometrical curvature also exists perforce in radially symmetric geometries, where $U(\vec r)=U(r)$. We consider a two-dimensional arena where an annular bulk is immured on both the inside and outside. This allows us to investigate more methodically how the active pressure depends on the magnitude and sign of the curvature.

Transforming the FPE into polar coordinates, we may find the recurrence relation for the steady-state density's moments, analogous to equation~(\ref{eqn:GeneralMoments1d}) in section~\ref{sec:moments1d} (see appendix~\ref{app:polarFPE} for the radially-symmetric expression). Let $\psi$ denote the angle between a particle's position vector and its propulsion force ($\vec r\cdot\vec \eta=r\eta\cos\psi$); then
\begin{align}
	\left<\eta\cos\psi\right>(r) = -f(r),
		\label{eqn:FPE_m1_polar}
\end{align}
which, similarly to equation~(\ref{eqn:FPE_m1}), means the radial component of the propulsion force must be balanced on average by the radial potential force. The second moment gives
\begin{align}
	\begin{split}
		0 = &-\partial_r\left[\left(\left<\eta^2\cos^2\psi\right>-f^2\right)n\right] +\\
			&\quad -\frac{1}{r}\left(\left<\eta^2\cos^2\psi\right>-\left<\eta^2\sin^2\psi\right>-f^2\right)n + \\
			&\quad +\frac{1}{\alpha}fn,
	\end{split}
	\label{eqn:FPE_polar_m11}
\end{align}
where all quantities are $r$-dependent. The Jacobian factor on the second line of equation~(\ref{eqn:FPE_polar_m11}) cannot be written as a total derivative with respect to $r$; thus there is no equation of state (see section~\ref{sec:pressure}). Nevertheless, since only the radial coordinate appears in equation~(\ref{eqn:FPE_polar_m11}), the integral to find the pressure exerted on a wall is insensitive to the path taken. It is natural to define $P=-\int f(r)n(r)\upd r$, and consequently
\begin{align}
	\begin{split}
		P = &\, \alpha \left<\eta^2\cos^2\psi\right> n |^\text{bottom of wall} +\\
			&- \alpha\int_\text{bottom of wall}^\text{top of wall}\frac{1}{r^\prime}\left(\left<\eta^2\cos^2\psi\right>+\right.\\
			&\left.\qquad\qquad\qquad\qquad-\left<\eta^2\sin^2\psi\right>-f^2\right)n \upd r^\prime,
	\end{split}
		\label{eqn:FPE_polar_P}
\end{align}
where ``bottom of wall'' could be anywhere in the bulk and ``top of wall'' may be at $r=0$ for the inner wall, or $r=\infty$ for the outer wall. Here we see explicitly the dependence of the pressure on statistical properties of particles in the wall region. However, when the bulk is large compared to the correlation length $\sqrt\alpha$, the first term in equation~(\ref{eqn:FPE_polar_P}) dominates, restoring an approximate equation of state. Furthermore, when the bulk region extends to $R \gg 1$ the wall curvature is small and the density resembles that for flat walls.

\begin{figure}[ht]
	\centering
	\includegraphics[width=\wid\textwidth]{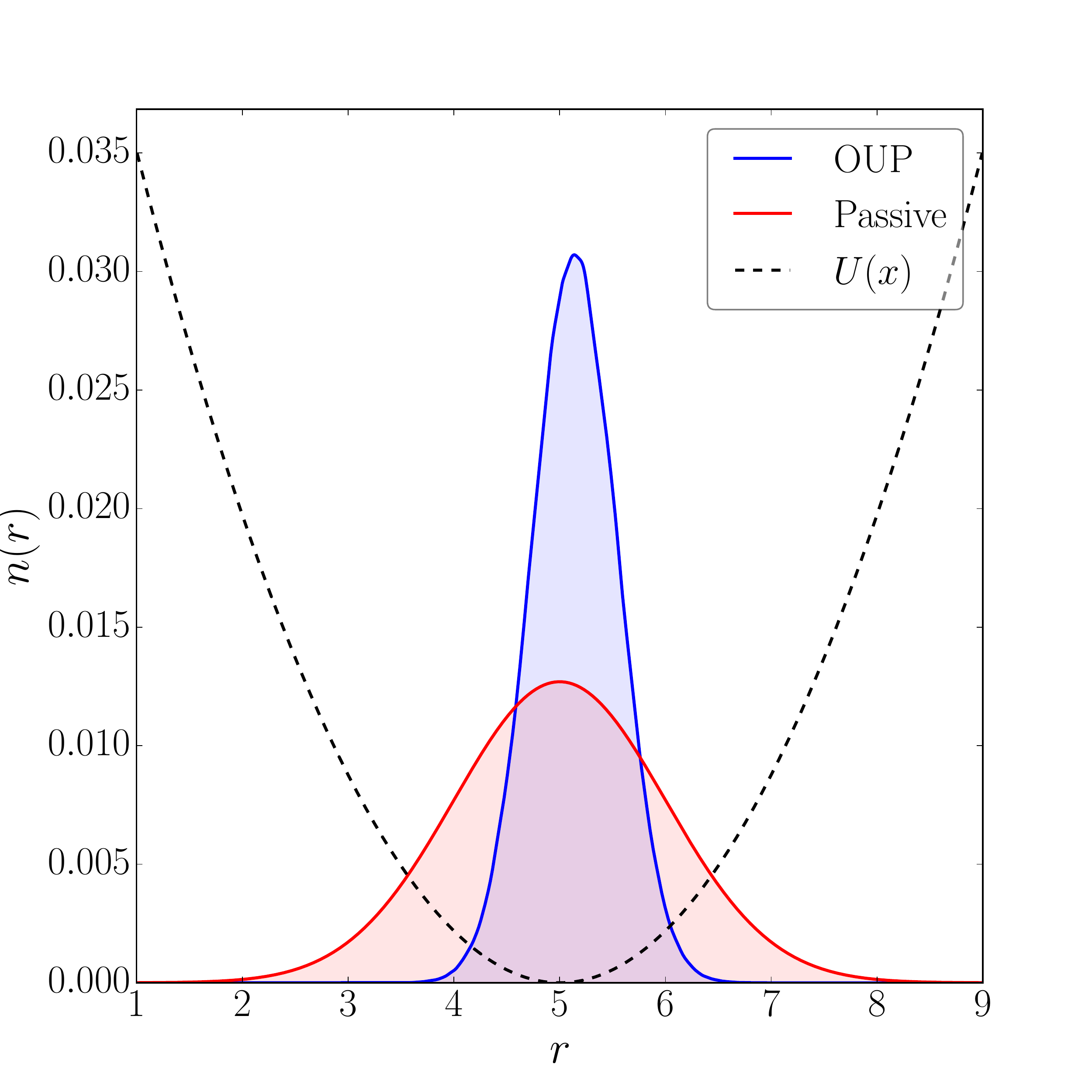}
	\caption{Spatial density $n(r)$ for annular geometry, with $U(r)=\frac{1}{2}(r-5)^2$ and $\alpha=5$. The OUP density is more sharply peaked than the passive particle density, and the average OUP position is offset from the bulk in the $+r$-direction, signalling a tendency for OUPs to collect in geometrically concave wall regions.}
	\label{fig:Q_POL_DR0}
\end{figure}

The discussion of curvature in the previous section suggests that the pressure on the outer wall must be larger than that on the inner wall (this can also be argued from the sign of the integral in equation~(\ref{eqn:FPE_polar_P})). This was verified numerically in \cite{CN1}, which considered the simplest potential with both signs of curvature, $U(r)=\frac{1}{2}(r-R)^2$, and found the pressure difference followed the difference in curvatures $\sim R^{-1}$. An example of a radial density profile is shown in Fig.~\ref{fig:Q_POL_DR0}.%

Recent work, based on the Unified Coloured Noise Approximation, has developed specific predictions for the density in radially-symmetric geometries \cite{Mag++15}. These nicely capture some qualitive features we observe, such as the shift of the maximum seen in Fig.~\ref{fig:Q_POL_DR0}. However, they also produce a number of inconsistencies, for instance predicting negative densities for radii $r/R<\alpha/(\alpha+1)$, and failing to capture accumulation effects around walls in one or more dimensions.

\begin{figure}[ht]
	\centering
	\includegraphics[width=\wid\textwidth]{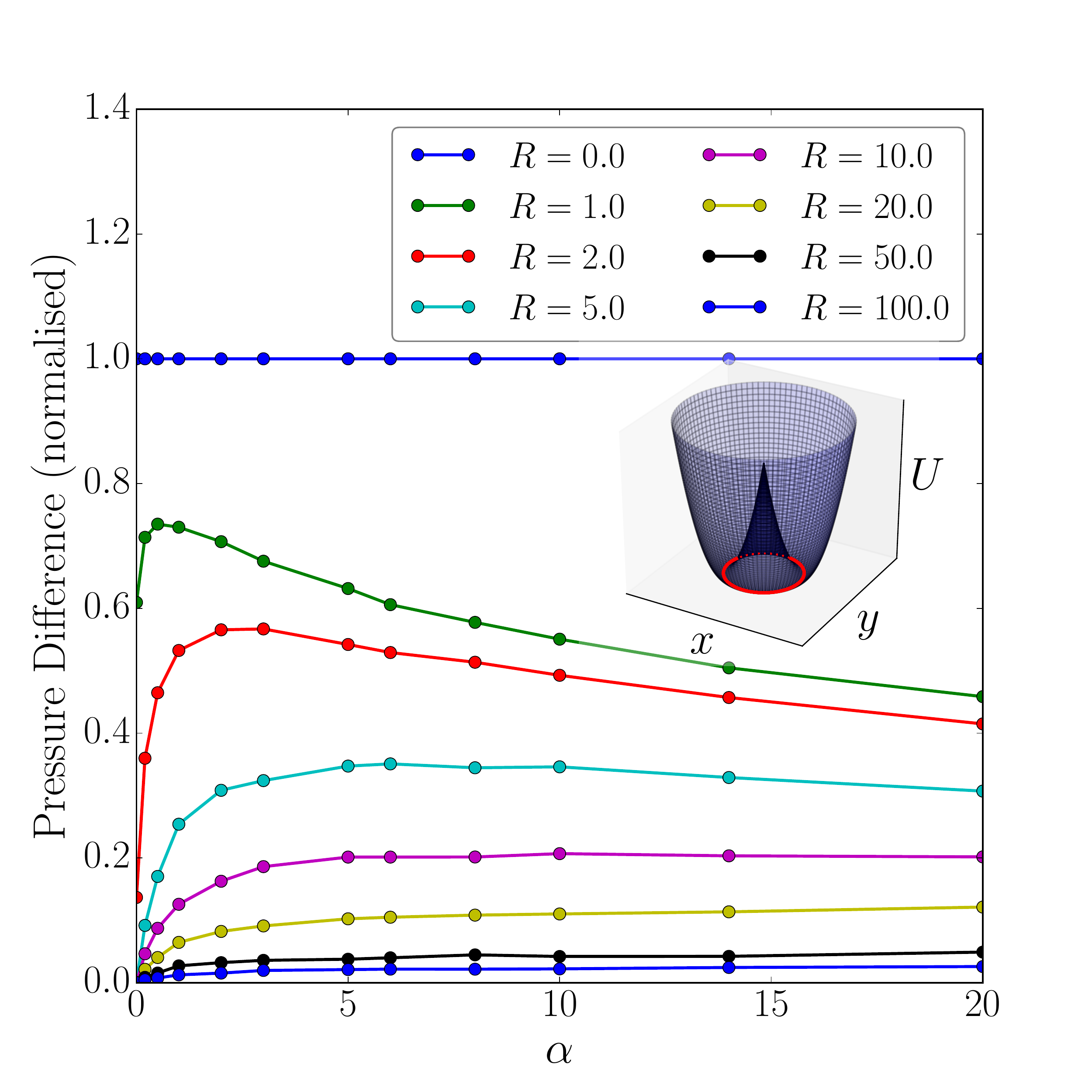}
	\caption{Pressure difference $(P_{\rm out}-P_{\rm in})/P^{\rm passive}_{\rm out}$ plotted against $\alpha$ for several bulk radii $R$. The potential is the radially symmetrical $U(r)=\frac{1}{2}(r-R)^2$, where the bulk has zero width. For a given $\alpha$, the pressure difference diminishes with $R$. Note that $\Delta P(\alpha=0)\neq0$ when $R$ is small enough that the inner wall is substantially penetrable. Note also that $\Delta P$ continues to increase slowly with $\alpha$ for $R\geq10$.}
	\label{fig:DPAS_POL_DR0}
\end{figure}

We now consider another aspect of this annular geometry, namely the role of persistence in OUPs' pressure with curved potentials. To this end, the pressure difference is plotted against $\alpha$ in Fig.~\ref{fig:DPAS_POL_DR0}. A number of features stand out. Firstly, when $R=0$ and the potential is a simple quadratic, $U=\frac{1}{2}r^2$, the pressure on the outer wall is $\alpha$-independent. This is a mathematically trivial fact in two dimensions.

The non-monotonicity in the pressure difference speaks to two competing regimes for the curvature.
At low $\alpha$, the relevant parameter is the persistence length times the curvature, since only particles with sufficiently high persistence can tell the geometry is curved.%
In the opposite limit, when $\alpha$ is large and the penetration is correspondingly small, the pressure on either wall and hence their difference must likewise become small. For a given $R$, these two effects balance at an $\alpha^\ast$ which can be roughly estimated from $\Delta P|_\text{small $\alpha$} \sim \frac{\sqrt\alpha}{R}$ {and} $\Delta P|_\text{large $\alpha$}\sim \frac{1}{\sqrt{\alpha+1}}$ which together imply the observed $\alpha^\ast\sim R$.

\begin{figure}[ht]
	\centering
	\includegraphics[width=\wid\textwidth]{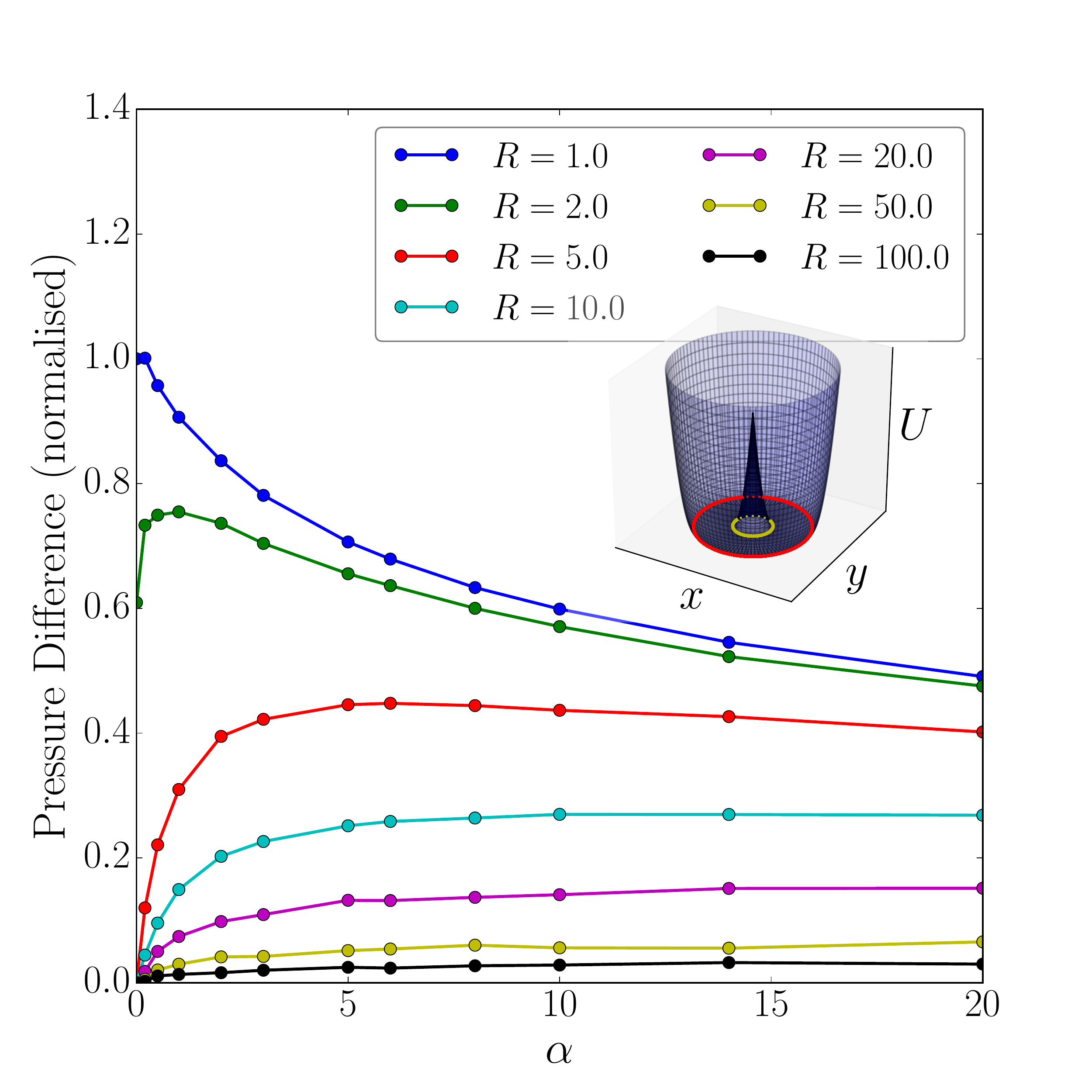}
	\caption{Analogous to Fig.~\ref{fig:DPAS_POL_DR0}, but where the bulk has finite width.}
	\label{fig:DPAR_POL_DR1}
\end{figure}

When the inner and outer walls differ in both curvature sign \textit{and} magnitude, as in the case of annular geometry with a \textit{finite} bulk size, these phenomena become somewhat distorted. Notation-wise, let the foot of the outer wall be at radial coordinate $R$ as before, and the foot of the inner wall be at coordinate $S$, such that $U(S\leq r\leq R)=0$. The pressure difference between outer and inner walls for a bulk of size 1 (that is, $R-S=1$) is presented in Fig.~\ref{fig:DPAR_POL_DR1}, and resembles in several respects the $R=S$ data from Fig.~\ref{fig:DPAS_POL_DR0}. However, we observe two small differences.

The first difference is that, at high $\alpha$, the existence of the finite bulk will raise the pressure on each wall for a given $\alpha$ and $R$. This is consistent with the findings of section~\ref{sec:CAR_DL}, where a larger bulk allows $\left<\vec\eta^2\right>$ to recover from its wall-depleted value. The outcome is to make the slope of $P(\alpha)$ shallower, as seen by comparing figures~\ref{fig:DPAS_POL_DR0} and~\ref{fig:DPAR_POL_DR1}.

The second effect, which manifests at low $\alpha$, is more interesting as it arises from the curvature difference. The inner wall has higher curvature than the outer wall ($1/S>1/R$), and as $\alpha$ increases from zero (and the OUPs acquire finite persistence length), its curvature will be felt first. Hence at low $\alpha$, the curves for a given $S$ match better than the curves for a given $R$.
(This can be seen by comparing the $R=0$ and $1$ curves from Fig.~\ref{fig:DPAS_POL_DR0} with the $R=1$ and $2$ curves of Fig.~\ref{fig:DPAR_POL_DR1}.)
The curvature of the outer wall only becomes relevant at higher $\alpha$.

Taking both these effects together, we see in Fig.~\ref{fig:DPAR_POL_DR1} that the peaks marking the crossover between curvature-dominated and flat wall behaviour are broadened compared to the zero-bulk case.

\section{Conclusion}
\label{sec:conclusion}

In this paper we sought to develop intuition for how the active force which propels Ornstein--Uhlenbeck Particles (OUPs) becomes depleted through  interaction with potentials, and how this manifests at long range due to OUPs' memory. These findings were employed to design several situations where obvious non-equilibrium effects would arise, the details of which were explored through numerical simulation.

In one spatial dimension, an asymmetrical arrangement of bulk regions (where there is zero external force) resulted in a net active pressure. These results appear qualitively similar to previous studies of active Brownian particles (ABPs) interacting with hard walls \cite{Sol++15}; however, the physics of the situation is somewhat different, as we demonstrate that neither alignment interactions nor long-range interactions are necessary for net active forces to develop. Thus the appearance of non-equilibrium behaviour is controlled not so much by the specific interactions between particles and walls, but by the regions where no interaction takes place.

In two dimensions, the role of spatial curvature on OUP behaviour was investigated numerically in two scenarios. First we explored the development of inhomogeneities in density along the contours of a soft wall, and connected them to an ad hoc definition of pressure. While our results were broadly in line with previous work for other swimmer models (most notably \cite{Nik++16}), we also observed differences such as a departure from the density-curvature relationship predicted in the high-persistence limit \cite{FBH15}.

The second test of curvature considered a radially-symmetric geometry, building on the work of \cite{CN1}. Having showed that no general equation of state exists for this system, we argued that competition between the active persistence length-scale and the curvature length-scale leads to non-monotonic pressure differences between convex and concave walls.

As a further remark, the annular geometry just discussed is somewhat reminiscent of a cylindrical rheometer. Our findings about how OUPs' stress responds to curvature indicate that care will have to be taken to properly define the viscoelastic moduli which are measured in experiments involving active matter. In particular, shearing with the inner cylinder or the outer cylinder will give different results.

\acknowledgements This work was supported primarily by the MRSEC Program of the National Science Foundation under Award Number DMR-1420073.  We benefited from collaboration with J.-F.Joanny on a number of related topics.  AYG acknowledges useful discussions with M.Kardar.



\begin{thebibliography}{10}

\bibitem{Mar++13}
M.~C. Marchetti, J.-F. Joanny, S.~Ramaswamy, T.~B. Liverpool, J.~Prost,
  M.~M.~Rao, and R.~A. Simha.
\newblock Hydrodynamics of soft active matter.
\newblock {\em Reviews of Modern Physics}, 85:1143--1189, July 2013.

\bibitem{Lee13}
C.~F. Lee.
\newblock Active particles under confinement: aggregation at the wall and
  gradient formation inside a channel.
\newblock {\em New Journal of Physics}, 15(5):055007, 2013.

\bibitem{E+G13}
J.~Elgeti and G.~Gompper.
\newblock Wall accumulation of self-propelled spheres.
\newblock {\em EPL (Europhysics Letters)}, 101:48003, February 2013.

\bibitem{YMM14}
X.~X.~Yang, L.~M. Manning, and C.~M. Marchetti.
\newblock Aggregation and segregation of confined active particles.
\newblock {\em Soft Matter}, 10:6477--6484, 2014.

\bibitem{Sol++15}
A.~P. Solon, Y.~Fily, A.~Baskaran, M.~E. Cates, Y.~Kafri, M.~Kardar, and
  J.~Tailleur.
\newblock Pressure is not a state function for generic active fluids.
\newblock {\em Nature Physics}, 11:673--678, August 2015.

\bibitem{T+C08}
J.~Tailleur and M.-E. Cates.
\newblock Statistical mechanics of interacting run-and-tumble bacteria.
\newblock {\em Physical Review Letters}, 100(21):218103, May 2008.

\bibitem{C+T15}
M.-E. Cates and J.~Tailleur.
\newblock Motility-induced phase separation.
\newblock {\em Annual Review of Condensed Matter Physics}, 6:219--244, March
  2015.

\bibitem{Mar++16}
U.~M.~B. Marconi, N.~Gnan, M.~Paoluzzi, C.~Maggi, and R.~{di Leonardo}.
\newblock Velocity distribution in active particles systems.
\newblock {\em Scientific Reports}, 6:23297, March 2016.

\bibitem{Sol++15--P-decomposition}
A.~P. Solon, J.~Stenhammar, R.~Wittkowski, M.~Kardar, Y.~Kafri, M.~E. Cates,
  and J.~Tailleur.
\newblock Pressure and phase equilibria in interacting active brownian spheres.
\newblock {\em Physical Review Letters}, 114(19):198301, May 2015.

\bibitem{HDL89}
P.~S. Hagan, C.~R. Doering, and C.~D. Levermore.
\newblock The distribution of exit times for weakly colored noise.
\newblock {\em Journal of Statistical Physics}, 54:1321--1352, March 1989.

\bibitem{DHL87}
C.~R. Doering, P.~S. Hagan, and C.~D. Levermore.
\newblock Bistability driven by weakly colored gaussian noise: The
  fokker-planck boundary layer and mean first-passage times.
\newblock {\em Physical Review Letters}, 59:2129--2132, November 1987.

\bibitem{Luc05}
J.~{\L}uczka.
\newblock Non-markovian stochastic processes: Colored noise.
\newblock {\em Chaos}, 15(2):026107, June 2005.

\bibitem{J+H87}
P.~Jung and P.~Hanggi.
\newblock Dynamical systems -- a unified colored-noise approximation.
\newblock {\em Phys. Rev. A}, 35:4464--4466, May 1987.

\bibitem{KML14}
N.~{Koumakis}, C.~{Maggi}, and R.~{Di Leonardo}.
\newblock {Directed transport of active particles over asymmetric energy
  barriers}.
\newblock {\em Soft Matter}, 10:5695, July 2014.

\bibitem{M+M15}
U.~M.~B. Marconi and C.~Maggi.
\newblock Towards a statistical mechanical theory of active fluids.
\newblock {\em Soft Matter}, 11:8768--8781, 2015.

\bibitem{Mag++15}
C.~Maggi, U.~M.~B. Marconi, N.~Gnan, and R.~{di Leonardo}.
\newblock Multidimensional stationary probability distribution for interacting
  active particles.
\newblock {\em Scientific Reports}, 5:10742, May 2015.

\bibitem{CN1}
C.~Sandford, A.Y. Grosberg, and J.-F. Joanny.
\newblock Pressure and flow of exponentially self-correlated active particles.
\newblock {\em Submitted to Physical Review E}, 2017.

\bibitem{vanKampen}
N.G. Van~Kampen.
\newblock {\em Stochastic Processes in Physics and Chemistry}.
\newblock North-Holland Personal Library. 2011.

\bibitem{Sza14}
G.~{Szamel}.
\newblock {Self-propelled particle in an external potential: Existence of an
  effective temperature}.
\newblock {\em \pre}, 90(1):012111, July 2014.

\bibitem{Fod++16}
{\'E}.~Fodor, C.~Nardini, M.-E. Cates, J.~Tailleur, P.~Visco, and F.~{van
  Wijland}.
\newblock How far from equilibrium is active matter?
\newblock {\em Physical Review Letters}, 117(3):038103, July 2016.

\bibitem{R+Z03}
R.~{Rzehak} and W.~{Zimmermann}.
\newblock {Inertial effects in Brownian motion of a trapped particle in shear
  flow}.
\newblock {\em Physica A Statistical Mechanics and its Applications},
  324:495--508, June 2003.

\bibitem{BrownianVertex_3}
Henrique~W. Moyses, Ross~O. Bauer, Alexander~Y. Grosberg, and David~G. Grier.
\newblock Perturbative theory for brownian vortexes.
\newblock {\em Phys. Rev. E}, 91:062144, Jun 2015.

\bibitem{MURA}
G. Falasco, F.Baldovin, K. Kroy, and M. Baiesi
\newblock Mesoscopic virial equation for
nonequilibrium statistical mechanics
\newblock {\em New Journal of Physics}, 18(9): 093043, 2016.

\bibitem{SFG17}
Stefano Steffenoni, Gianmaria Falasco, and Klaus Kroy.
\newblock Microscopic derivation of the hydrodynamics of
  active-brownian-particle suspensions.
\newblock {\em Phys. Rev. E}, 95:052142, May 2017.

\bibitem{LL_volV}
Lev.~D. Landau, and Evgeny~M.Lifshitz
\newblock{Statistical Physics Part 1 (Course of Theoretical Physics, Volume 5)}
\newblock{Butterworth-Heinemann; 3 edition, 1980}

\bibitem{S+J16}
T.~Speck and R.~L. Jack.
\newblock Ideal bulk pressure of active brownian particles.
\newblock {\em Phys. Rev. E}, 93(6):062605, June 2016.

\bibitem{J+B16}
M.~Joyeux and E.~Bertin.
\newblock Pressure of a gas of underdamped active dumbbells.
\newblock {\em Phys. Rev. E}, 93(3):032605, March 2016.

\bibitem{WWG15}
R.~G. {Winkler}, A.~{Wysocki}, and G.~{Gompper}.
\newblock {Virial pressure in systems of spherical active Brownian particles}.
\newblock {\em Soft Matter}, 11:6680--6691, 2015.

\bibitem{MMM16}
U.~{Marini Bettolo Marconi}, C.~{Maggi}, and S.~{Melchionna}.
\newblock {Pressure and surface tension of an active simple liquid: a
  comparison between kinetic, mechanical and free-energy based approaches}.
\newblock {\em Soft Matter}, 12:5727--5738, 2016.

\bibitem{FBH14}
Y.~Fily, A.~Baskaran, and M.~F. Hagan.
\newblock Dynamics of self-propelled particles under strong confinement.
\newblock {\em Soft Matter}, 10:5609--5617, 2014.

\bibitem{FBH15}
Y.~Fily, A.~Baskaran, and M.~F. Hagan.
\newblock Dynamics and density distribution of strongly confined noninteracting
  nonaligning self-propelled particles in a nonconvex boundary.
\newblock {\em Phys. Rev. E}, 91:012125, January 2015.

\bibitem{S+L15}
F.~Smallenburg and H.~L{\"o}wen.
\newblock Swim pressure on walls with curves and corners.
\newblock {\em Phys. Rev. E}, 92(3):032304, September 2015.

\bibitem{Nik++16}
N.~Nikola, A.~P. Solon, Y.~Kafri, M.~Kardar, J.~Tailleur, and R.~Voituriez.
\newblock Active particles with soft and curved walls: Equation of state,
  ratchets, and instabilities.
\newblock {\em Physical Review Letters}, 117(9):098001, August 2016.

\bibitem{Gal++07}
P.~Galajda, J.~Keymer, P.~M. Chaikin, and R.~Austin.
\newblock A wall of funnels concentrates swimming bacteria.
\newblock {\em J. Bacteriol.}, 189:8704--8707, December 2007.

\bibitem{PhysRevE.89.032720}
A.~Guidobaldi, Y.~Jeyaram, I.~Berdakin, V.~V. Moshchalkov, C.~A. Condat, V.~I.
  Marconi, L.~Giojalas, and A.~V. Silhanek.
\newblock Geometrical guidance and trapping transition of human sperm cells.
\newblock {\em Phys. Rev. E}, 89:032720, Mar 2014.

\end{thebibliography}


\appendix


\section{A General Correlation Function}
\label{app:GeneralCorrelation}

Consider an overdamped particle driven by a stochastic propulsion force $\vec\Xi(t)$ of zero mean and an arbitrary self-correlation function $\phi(t)$. Mathematically, we have
\begin{align}
	\dot{\vec x} = \vec f(\vec x)+ \vec{\Xi}(t)
\end{align}
\begin{align}
	\left< \Xi_i(t)\right>=0 \qquad \left<\Xi_i(t)\Xi_j(t^\prime)\right>=\delta_{ij}\phi(|t-t^\prime|),
\end{align}
where vector components are labelled by the indices $i,j$. This propulsion force can be constructed from a weighted sum of \textit{exponentially} correlated stochastic forces $\vec\eta_\alpha$, which have dimensionless correlation time $\alpha$, and whose components obey the equation
\begin{align}
	\alpha\dot\eta_{\alpha,i} = -\eta_{\alpha,i} + \xi_i^{\alpha}(t),
\end{align}
where $\xi_i^{\alpha}(t)$ is a Gaussian white noise with $\left<\xi_i^{\alpha}(t)\xi_j^{\alpha^\prime}(t^\prime)\right>=2\delta_{i,j}\delta_{\alpha,\alpha^\prime}\delta(t-t^\prime)$. Let the weights in the sum be called $w_\alpha$, such that
\begin{align}
	\Xi_i(t) = \int_0^\infty w_\alpha \eta_{\alpha,i}(t) \upd\alpha.
		\label{eqn:GeneralCorrelationWeightedSum}
\end{align}

The relationship between the weights $w_\alpha$ and the correlation function $\phi(t)$ is as follows. From equation~(\ref{eqn:GeneralCorrelationWeightedSum}), we compute the correlation function
\begin{align}
	\left< \Xi_i(t)\Xi_j(t^\prime)\right> &= \left< \int_0^\infty \upd\alpha \int_0^\infty \upd\alpha^\prime w_\alpha w_{\alpha^\prime} \eta_{\alpha,i}(t) \eta_{\alpha^\prime,j}(t^\prime) \right>
		\nonumber
	\intertext{The correlator in the left is the known function $\phi(|t-t^\prime|)$, while on the right the only random variables are the exponentially-correlated components of $\vec\eta_\alpha$. Thus,}
	\begin{split}
		\phi(t) &= \int_0^\infty \upd\alpha \int_0^\infty \upd\alpha^\prime\; w_\alpha w_{\alpha^\prime}\times\\
			&\qquad\qquad\qquad\times\delta(\alpha-\alpha^\prime)\frac{1}{\alpha}\exp\left[-{t}/{\alpha}\right]
				\nonumber
	\end{split}\\
	& = \int_0^\infty \upd\alpha\;  w_\alpha^2 \frac{1}{\alpha}\exp\left[-{t/\alpha}\right]
		\label{eqn:GeneralCorrelationLaplaceTransformA}
\end{align}
{Recognising equation~(\ref{eqn:GeneralCorrelationLaplaceTransformA}) as a Laplace transform between conjugate variables $t$ and $\frac{1}{\alpha}$, we can write the weights $w_\alpha$ as the inverse transform}
\begin{align}
	w_\alpha = \sqrt{\alpha \mathcal L^{-1}\left[\phi(t)\right](\alpha)}.
\end{align}
Provided the inverse transform $\mathcal L^{-1}\left[\phi(t)\right](\alpha)$ exists, an arbitrarily correlated noise $\vec\Xi(t)$ can therefore be constructed from exponentially-correlated noise $\vec\eta(t)$.




\section{The Influence of an Interior Wall on Propulsion Distribution and Net Force}
\label{app:filter}
\label{app:CAR_ML_netforce}

\begin{figure}[ht]
	\centering
	\includegraphics[width=\wid\textwidth]{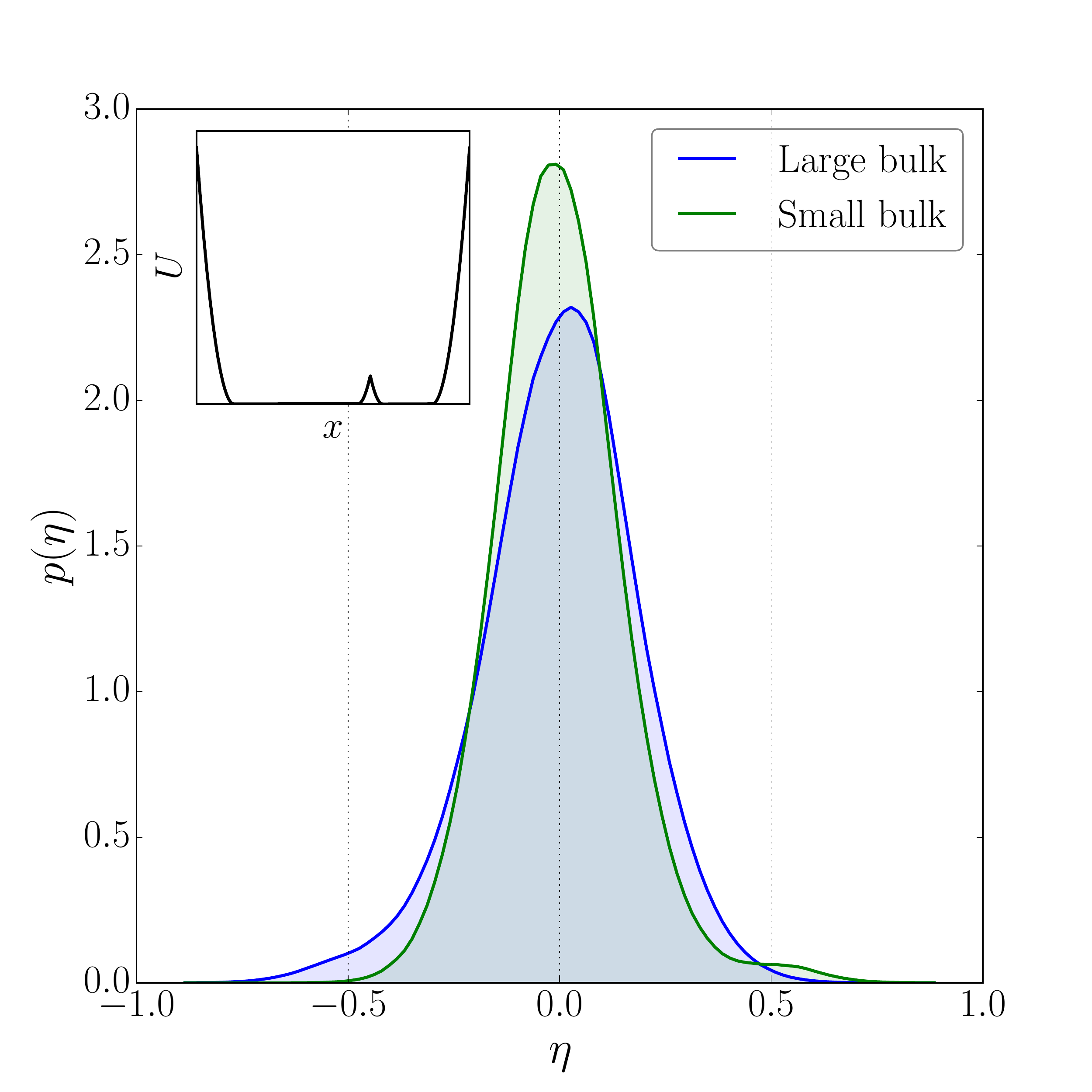}
	\caption{Distribution of $\eta$ (normalised) in a large bulk and a small bulk. The potential at the cusp of the interior wall is $0.125$ (in units of $T$) and the force magnitude is $0.5$ (in units of $\sqrt{Tk}$).}
	\label{fig:filter}
\end{figure}

In section~\ref{sec:CAR_ML}, we considered a scenario with an OUP confined in a volume which featured a small (that is, penetrable) potential barrier. The distribution of $\eta$ in the bulks on either side of this barrier was already shown in the inset of Fig.~\ref{fig:PA_CAR_ML}. In Fig.~\ref{fig:filter} we show another example for an even smaller interior wall.

For very large bulks, the two distributions in Fig.~\ref{fig:filter} would be Gaussian, but here we see two non-Gaussian features. The first is for the small bulk around $\eta=+0.5$, where particles from the large bulk are able to overcome the barrier (whose maximum force is $0.5$); the barrier is here acting as a kind of filter for high-$\eta$ particles. Since these particles will go on to interact with the confining wall which diminishes their $\eta$, there is no analogous limb around $\eta=-0.5$. The second, related feature is the compensating distortion of the distribution in the large bulk, which preferentially retains its small-positive-$\eta$ and large-negative-$\eta$ particles.

\begin{figure}[ht]
	\centering
	\includegraphics[width=\wid\textwidth]{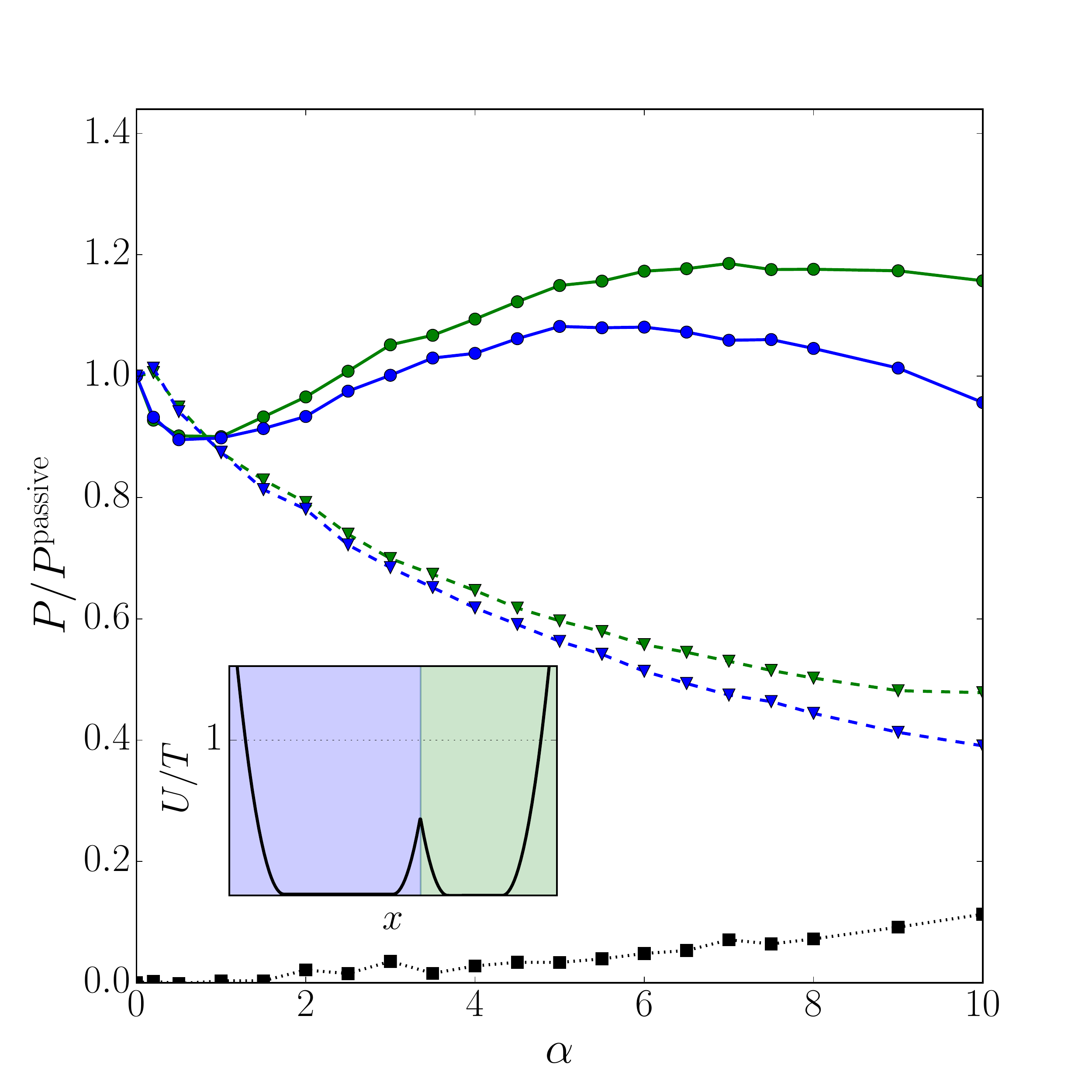}
	\caption{Pressure on either side of a penetrable inner wall (as in Fig.~\ref{fig:PA_CAR_ML}) along with the pressure on the confining walls (dashed triangles). The net pressure in the $-x$ direction is plotted in black squares connected by dots. This is the same data as for Fig.~\ref{fig:PA_CAR_ML} in the main text.}
	\label{fig:PAall_CAR_ML}
\end{figure}

Fig.~\ref{fig:PAall_CAR_ML} shows the pressure on all four walls (that is, the two sides of the interior wall and the two confining walls), along with the net pressure which points in the $-x$ direction, and would translate the volume at a constant rate if it were free to move.

\section{Pressure on a Wall which is Periodic in One Cartesian Dimension}
\label{app:pressure_periodic}

Consider a potential which is periodic in one Cartesian dimension. For the sake of simplicity and applicability to the results presented in section~\ref{sec:CAR_UL}, we restrict attention to two dimensions, with $U(x,y)=U(x,y+\lambda)$. (The procedure and result works for higher dimensions, but the notation becomes that bit more cumbersome.) Evaluating equation~(\ref{eqn:FPE_m1_ind}) in the bulk ($f_i=0$) and expanding the sum,
\begin{align}
	0 &= \partial_x\left[\left<\eta_x^2\right>(x,y) n(x,y)\right] + \partial_y\left[\left<\eta_x\eta_y\right>(x,y)n(x,y)\right]
		\nonumber
\end{align}
{Integrating this over $x$ within the bulk, and dropping coordinate dependencies,}
\begin{align}
	\left<\eta_x^2\right> n|^{\rm bulk} &= c_x - \int^{\rm bulk} \partial_y\left[\left<\eta_x\eta_y\right> n \right] \upd x,
		\label{eqn:etaxQ2_bulk_2D}
\end{align}
where ``$|^\text{bulk}$'' means evaluated in the bulk, the integral is an indefinite integral anywhere within the bulk, and $c_x$ is an integration constant. If there is no variation in the $y$-direction (ie the wall is flat) we are back to the quasi-1D scenario explored in section~\ref{sec:moments1d}: the integral in equation~(\ref{eqn:etaxQ2_bulk_2D}) is zero, and using equation~(\ref{eqn:PBC}) we can identify $c_x=\frac{1}{\alpha}P_\text{flat}$.

To find the ``local pressure'' on a confining wall defined in equation~(\ref{eqn:Py}), we integrate the $x$-component of the generally valid equation~(\ref{eqn:FPE_m1_ind}) over $x$ from somewhere in the bulk to $\infty$:
\begin{align}
	\frac{1}{\alpha} P_x(y) &= \left<\eta_x^2\right> n|^\text{bulk} - \partial_y\left[\int_\text{bulk}^\infty \left(\left<\eta_x\eta_y\right>-f_x f_y\right) n \upd x\right],
		\nonumber
\end{align}
{Combining with equation~(\ref{eqn:etaxQ2_bulk_2D}),}
\begin{align}
	\begin{split}
		\frac{1}{\alpha} P_x(y) = c_x - \partial_y&\left[\int^\text{bulk}\left<\eta_x\eta_y\right> n \upd x +\right.\\
			&\left. +\int_\text{bulk}^\infty \left(\left<\eta_x\eta_y\right>-f_x f_y\right) n \upd x\right].
	\end{split}
	\label{eqn:Py_2D}
\end{align}
The total pressure (in the $x$-direction) on a complete period of the wall is denoted $P_x$ and is equal to
\begin{align}
	P_x &= \frac{1}{\lambda} \int_0^\lambda P_x(y) \upd y.
	\intertext{Substituting in equation~(\ref{eqn:Py_2D}), and using the periodicity,}
	P_x &= P_\text{flat}
\end{align}

When the wall is not symmetric in the $y$-direction, there will also be a net local pressure $P_y(x)$ \cite{Nik++16}. However the \textit{total} pressure in the $y$-direction for any $y$-periodic wall must be zero.

\section{Moments of Fokker--Planck Equation in a Radially Symmetrical Potential}
\label{app:polarFPE}

\begin{figure}[ht]
	\centering
	\includegraphics[width=\wid\textwidth]{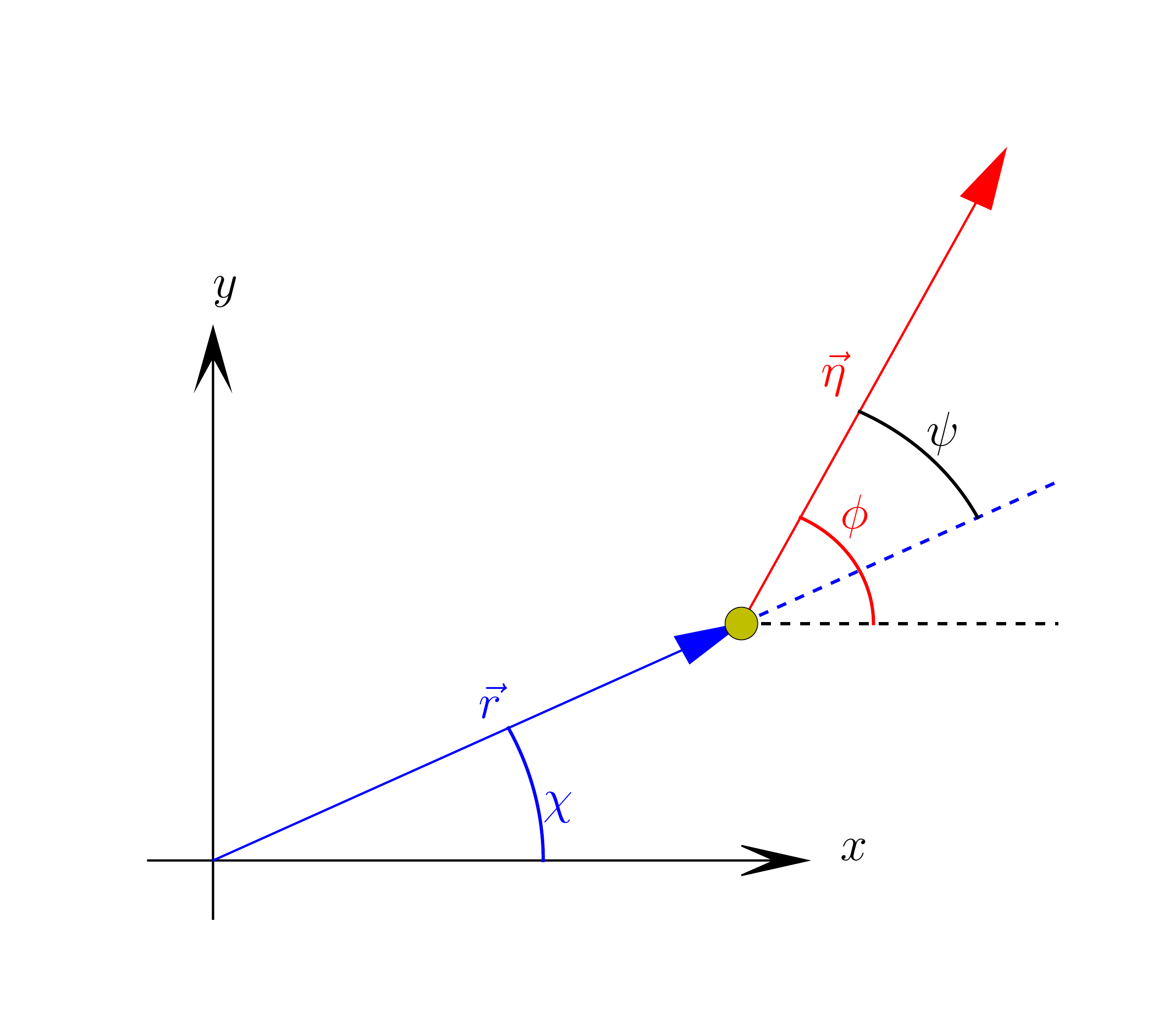}
	\caption{Coordinate system for the radially symmetric geometry.}
	\label{fig:Coords_POL}
\end{figure}

The vector form of the steady-state FPE, where the density $\rho=\rho(\vec r,\vec\eta)$, is
\begin{align}
	0 &= -\nabla_{\vec r}\cdot\left[\left(\vec\eta+\vec f(\vec r)\right)\rho\right] + \frac{1}{\alpha}\nabla_{\vec\eta}\cdot\left[\vec\eta\rho\right] + \frac{1}{\alpha^2}\nabla_{\vec\eta}^2\left[\rho\right].
		\label{eqn:FPE_vec}
\end{align}
Assuming we are in a {radially-symmetric} environment -- that is, the potential force $\vec f(\vec r)=f(r)\hat r$, it is reasonable to translate equation~(\ref{eqn:FPE_vec}) into polar coordinates (see Fig.~\ref{fig:Coords_POL} for coordinate system):
\begin{align}
	\vec r = \begin{pmatrix}x\\ y\end{pmatrix} = \begin{pmatrix}r\cos\chi\\ r\sin\chi\end{pmatrix} \qquad
	\vec\eta = \begin{pmatrix}\eta_x\\ \eta_y\end{pmatrix} = \begin{pmatrix}\eta\cos\phi\\ \eta\sin\phi\end{pmatrix}.
		\nonumber
\end{align}
And the angle
\begin{align}
	\psi\equiv\phi-\chi.
		\label{eqn:psi}
\end{align}
This represents angle between $\vec\eta$ and the radial direction to the particle -- that is, $\vec\eta\cdot\vec r=\eta r\cos\psi$.

Translating the vector derivatives in equation~(\ref{eqn:FPE_vec}) into our coordinate system,
\begin{align}\begin{split}
	0 = &-\partial_r\left[\left(\eta\cos\psi+f\right)\rho\right] - \frac{1}{r}\left(\eta\cos\psi+f\right)\rho +\\
		&- \frac{1}{r}\partial_\chi\left[(\eta\sin\psi)\rho\right] + \frac{1}{\alpha}\left( \partial_\eta\left[\eta\rho\right] + \rho \right) +\\
		&+ \frac{1}{\alpha^2}\left( \partial_\eta^2\left[\rho\right] + \frac{1}{\eta}\partial_\eta\left[\rho\right] + \frac{1}{\eta^2}\partial_\phi^2\left[\rho\right] \right),
			\label{eqn:FPE_polar_chiphipsi}
\end{split}\end{align}
where the first three terms come from spatial advection; the fourth term is the response of the random propulsion magnitude, $\eta$, to its ``restoring force''; and the remainder are diffusive spreading of the random propulsion.

To capitalise on symmetry, we write equation~(\ref{eqn:FPE_polar_chiphipsi}) in terms of the angle $\psi$ only:
\begin{align}\begin{split}
	0 = &-\partial_r\left[\left(\eta\cos\psi+f\right)\rho\right] - \frac{1}{r}\left(\eta\cos\psi+f\right)\rho + \\
		&+\frac{1}{r}\partial_\psi\left[\left(\eta\sin\psi\right)\rho\right] + \frac{1}{\alpha}\left(\partial_\eta\left[\eta\rho\right] + \rho \right) +\\
		&+ \frac{1}{\alpha^2}\left( \partial_\eta^2\left[\rho\right] + \frac{1}{\eta}\partial_\eta\left[\rho\right] + \frac{1}{\eta^2}\partial_\psi^2\left[\rho\right] \right).
			\label{eqn:FPE_polar_psi}
\end{split}\end{align}

Define the $r$-dependent moments as
\begin{align}
	\begin{split}
		\left<\eta^n\cos^m\right.\!&\left.\!\psi\right>(r)n(r) = \\
			& \int_0^{2\pi}\int_0^\infty \eta^n\cos^m\psi \,\rho(r,\eta,\psi) \,\eta\upd\eta\upd\psi,
	\end{split}
		\label{eqn:moment_definition}
\end{align}
we can integrate equation~(\ref{eqn:FPE_polar_psi}) over angle $\psi$ and noise magnitude $\eta$ to eventually arrive at an expression relating moments for any $n$ and $m$.
\begin{align}\begin{split}
	0 = &-\partial_r\left[\left(\left<\eta^{n+1}\cos^{m+1}\psi\right> + \left<\eta^{n}\cos^{m}\psi\right> f \right) n\right] \,+\\
		&-\frac{1}{r}\left(\left<\eta^{n+1}\cos^{m+1}\psi\right> + \left<\eta^{n}\cos^{m}\psi\right> f \right) n \,+\\
		&+\frac{1}{r}m\left(\left<\eta^{n+1}\cos^{m-1}\psi\right> - \left<\eta^{n+1}\cos^{m+1}\psi\right>\right) n \,+\\
		&-\frac{1}{\alpha}n\left<\eta^n\cos^{m}\psi\right> n \,+\\
		&+\frac{1}{\alpha^2}(n^2-m^2)\left<\eta^{n-2}\cos^{m}\psi\right> n \,+\\
		&+\frac{1}{\alpha^2}m(m-1)\left<\eta^{n-2}\cos^{m-2}\psi\right> n
			\label{eqn:moment_general}
\end{split}\end{align}

\section{Annular Geometry -- Pressure on Each Wall}
\label{app:annulus_individual_pressure}

\begin{figure}
	\centering
	\includegraphics[width=\wid\textwidth]{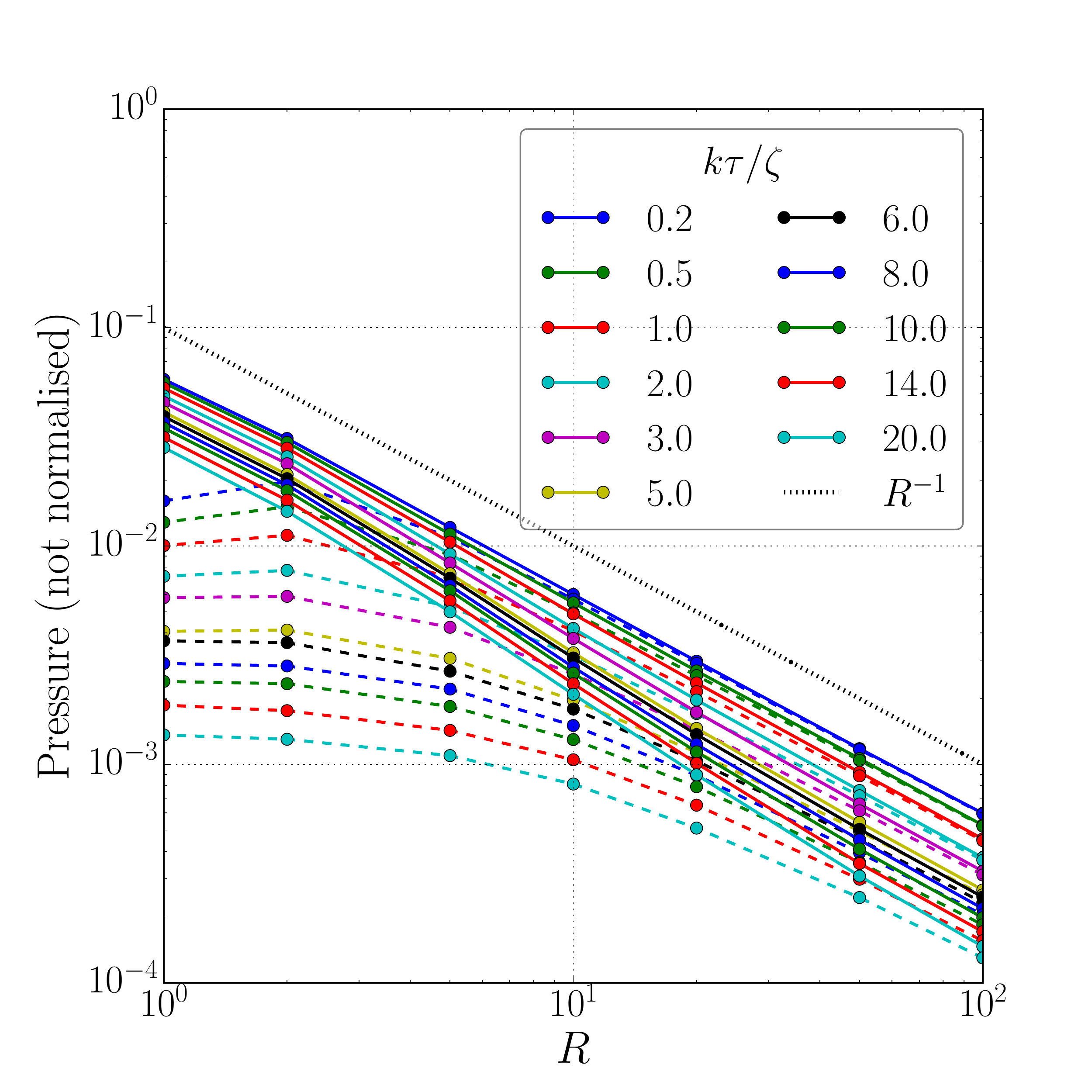}
	\caption{Pressure on the outer (solid lines) and inner (dashed) as a function of $R$ for several values of $\alpha$, in the annular geometry with zero bulk ($R=S$). The pressure is not normalised.}
	\label{fig:PQ_POL_DL_DR0}
\end{figure}

ABPs confined by hard ellipsoidal walls were found to exert a pressure which fell as $1/R$ in the high-persistence limit \cite{FBH14}. The same relation is seen in Fig.~\ref{fig:PQ_POL_DL_DR0} (solid lines) for OUPs interacting with soft circular walls, and is due simply to the dilution of density as the system growa.

The pressure exerted on the convex inner wall (dashed lines of corresponding colour) is more interesting, and starts to behave like $1/S$ only when the curvature is very small.

\section{Disc Geometry}
\label{app:disc}

\begin{figure}
	\centering
	\includegraphics[width=\wid\textwidth]{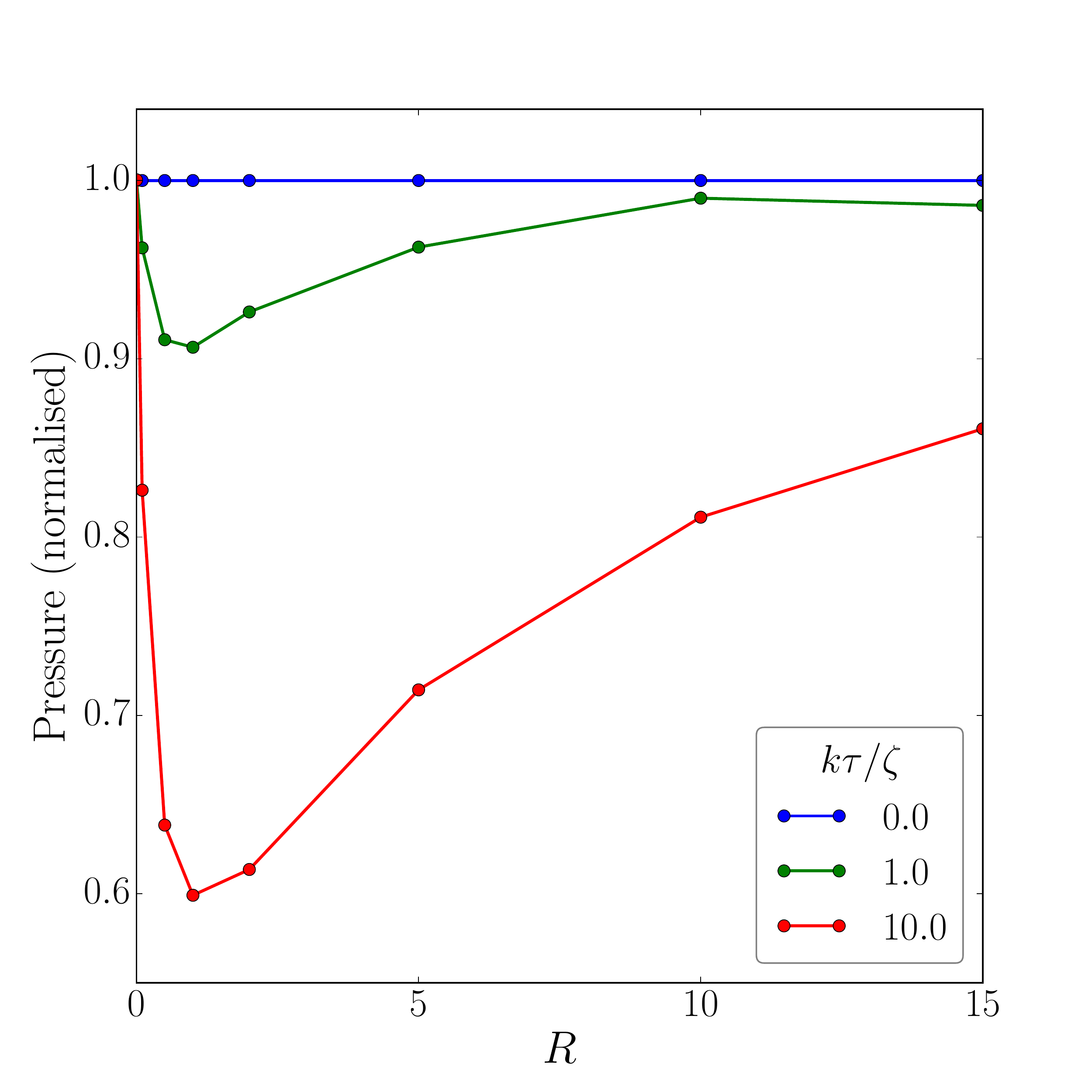}
	\caption{Pressure as a function of $R$ for several values of $\alpha$, in a disc geometry where there is no inner wall. The pressure is normalised by the passive value, and is higher for higher $\alpha\equiv k\tau/\zeta$ (for a fixed $R$).}
	\label{fig:PR_POL_L}
\end{figure}

Consider a radially-symmetric geometry where OUPs are confined by an outer wall but there is no inner wall (so the bulk is a disc rather than an annulus). For $\alpha=0$, we expect the pressure to equal the passive pressure for obvious reasons. When $R=0$, and the bulk is just a point at $r=0$, we also expect the pressure to equal the passive pressure, in the present case of a quadratic potential. Furthermore, equation~(\ref{eqn:FPE_polar_P}) for the pressure in a radial geometry tells us that when the bulk is very large, the pressure will equal the passive pressure once more. Thus we have $P\approx P^{\rm passive}$ for both low and high $R$, with some non-trivial interpolation in-between. This is shown in Fig.~\ref{fig:PR_POL_L}.

\section{Linear Potential}
\label{app:linpot}

\begin{figure}
	\centering
	\includegraphics[width=\wid\textwidth]{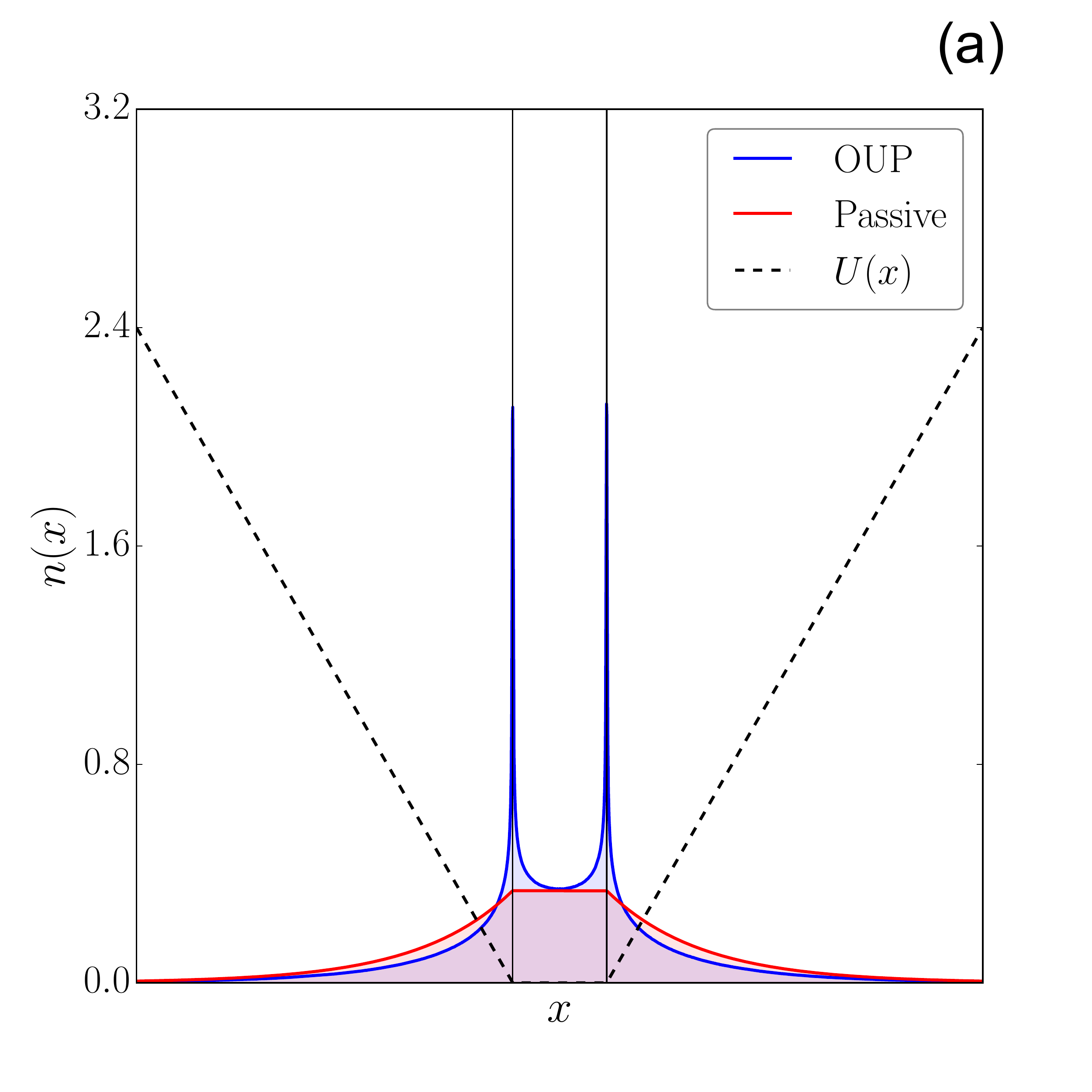}
	\includegraphics[width=\wid\textwidth]{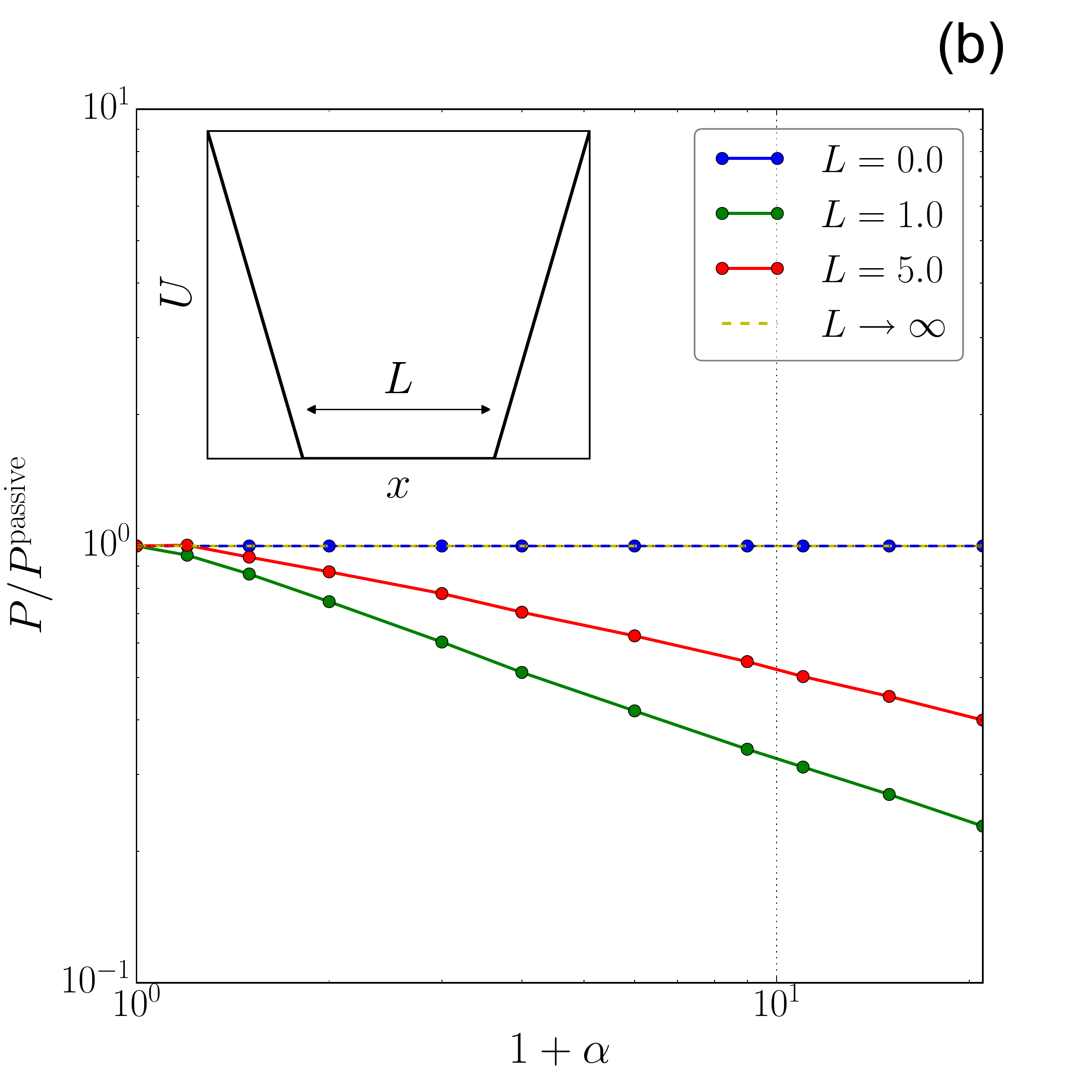}
	\caption{\textbf{(a)} Spatial density for an OUP confiend in a piecewise linear potential (sketched as dashed black lines).
	\textbf{(b)} Pressure as a function of $\alpha$ for the linear potential pictured in the inset. Note the $L\to\infty$ line conicides with the $L=0$ line.}
	\label{fig:PDF_CAR_DC}
	\label{fig:PA_CAR_DC}
\end{figure}

Though we mostly restricted attention to piece-wise quadratic potentials, the same qualitative behaviours are observed with other potentials. Here we consider the example of a piece-wise linear potential in 1D. The upper panel (a) of Fig.~\ref{fig:PDF_CAR_DC} shows the spatial density of an OUP confined in one dimension with finite bulk (compare with the blue curve in Fig.~\ref{fig:BC_CAR_DL_FB}), while the lower panel (b) plots the pressure exerted on the confining walls (compare with Fig.~2 in \cite{CN1}). Note that $P(\alpha;L=0)=1$ for the same reason as the $L=0$ line in Fig.~\ref{fig:DPAS_POL_DR0}.

\end{document}